\documentclass[11pt, a4paper,nofootinbib]{article}
\pdfoutput=1
\usepackage{jheppub}
\usepackage{amsfonts}
\usepackage{amsthm}
\usepackage{multirow}
\usepackage{graphicx}
\usepackage{harpoon}
\usepackage{tikz}
\begin{document}
\newcommand{\N}{\mathbb{N}}
\newcommand{\M}{\mathbb{M}}
\newcommand{\Z}{\mathbb{Z}}
\newcommand{\C}{\mathbb{C}}
\newcommand{\R}{\mathbb{R}}
\newcommand{\F}{\mathbb{F}}
\newcommand{\defineL}{\,\mathrel{\mathop:}=}
\newcommand{\defineR}{ =\mathrel{\mathop:}\,}
\newcommand{\be}{\begin{equation}}
\newcommand{\ee}{\end{equation}}
\newcommand{\bi}{\begin{itemize}}
\newcommand{\ei}{\end{itemize}}
\newcommand{\with}{\qquad\text{with}\qquad}
\newcommand{\mand}{\qquad\text{and}\qquad}
\newcommand{\sep}{\; \text{,}\qquad}
\newcommand{\ssep}{\; \text{,}\quad}
\newcommand{\com}{\; \text{,}}
\newcommand{\pt}{\: \text{.}}
\newcommand{\pd}{\partial}
\newcommand{\pdo}{\overline{\partial}}
\newcommand{\ov}[1]{\overline{#1}}
\newcommand{\inv}[1]{\genfrac{}{}{}{0}{1}{#1}}
\newcommand{\tinv}[1]{\genfrac{}{}{}{1}{1}{#1}}
\newcommand{\ttinv}[1]{\genfrac{}{}{}{2}{1}{#1}}
\newcommand{\tttinv}[1]{\genfrac{}{}{}{3}{1}{#1}}
\newcommand{\abs}[1]{|#1|}
\newcommand{\B}[1]{\mathbf{#1}}
\newcommand{\op}[1]{\mathrm{#1}}
\newcommand{\bracket}[1]{\left( #1 \right)}
\newcommand{\bracketi}[1]{\bigl( #1 \bigr)}
\newcommand{\bracketii}[1]{\Bigl( #1 \Bigr)}
\newcommand{\bracketiii}[1]{\biggl( #1 \biggr)}
\newcommand{\bracketiv}[1]{\Biggl( #1 \Biggr)}
\newcommand{\bpm}{\begin{pmatrix}}
\newcommand{\epm}{\end{pmatrix}}
\newcommand{\bnm}{\begin{matrix}}
\newcommand{\enm}{\end{matrix}}
\newcommand{\bt}{\begin{tikzpicture}}
\newcommand{\et}{\end{tikzpicture}}
\newcommand{\w}{\wedge}
\newcommand{\tp}{\otimes}
\newcommand{\ds}{\oplus}
\newcommand{\no}{\nonumber}
\newcommand{\D}{\mathcal{D}}
\newcommand{\vt}{\theta}
\newcommand{\Dvt}{D_\theta}
\newcommand{\vect}[1]{\text{\overrightharp{\ensuremath{#1}}}}
\newcommand{\nr}{n_{\scriptscriptstyle R}}
\newcommand{\nn}{n_{\scriptscriptstyle NS}}
\newcommand{\nrr}{n_{\scriptscriptstyle R-R}}
\newcommand{\nrn}{n_{\scriptscriptstyle R-NS}}
\newcommand{\nnr}{n_{\scriptscriptstyle NS-R}}
\newcommand{\nnn}{n_{\scriptscriptstyle NS-NS}}
\newcommand{\tnr}{\tilde{n}_{\scriptscriptstyle R}}
\newcommand{\tnn}{\tilde{n}_{\scriptscriptstyle NS}}
\newcommand{\tnrr}{\tilde{n}_{\scriptscriptstyle R-R}}
\newcommand{\tnrn}{\tilde{n}_{\scriptscriptstyle R-NS}}
\newcommand{\tnnr}{\tilde{n}_{\scriptscriptstyle NS-R}}
\newcommand{\tnnn}{\tilde{n}_{\scriptscriptstyle NS-NS}}
\newcommand{\ti}[1]{\tilde{#1}}

\newcommand{\oF}{\mathcal{F}}
\newcommand{\oP}{\mathcal{P}}
\newcommand{\oQ}{\mathcal{Q}}
\newcommand{\oo}[2]{\sideset{_{#1}}{_{#2}}{\mathop{\circ}}}
\newcommand{\oxi}[1]{{\mathop{\xi}}_{#1}}
\newcommand{\ooP}[2]{\overset{\scriptscriptstyle P_\alpha}{\sideset{_{#1}}{_{#2}}{\mathop{\circ}}}}
\newcommand{\oxiP}[1]{\overset{\scriptscriptstyle P_\alpha}{\xi_{#1}}}
\newcommand{\oophi}[2]{\overset{\scriptscriptstyle \Phi_\alpha}{\sideset{_{#1}}{_{#2}}{\mathop{\circ}}}}
\newcommand{\oxiphi}[1]{\overset{\scriptscriptstyle \Phi_\alpha}{\xi_{#1}}}
\newcommand{\ooI}[2]{\overset{\scriptscriptstyle I_\alpha}{\sideset{_{#1}}{_{#2}}{\mathop{\circ}}}}
\newcommand{\oxiI}[1]{\overset{\scriptscriptstyle I_\alpha}{\xi_{#1}}}
\newcommand{\oobpz}[2]{\overset{\scriptscriptstyle \op{bpz}_\alpha}{\sideset{_{#1}}{_{#2}}{\mathop{\circ}}}}
\newcommand{\oxibpz}[1]{\overset{\scriptscriptstyle \op{bpz}_\alpha}{\xi_{#1}}}
\newcommand{\oosym}[2]{\overset{\scriptscriptstyle \omega_\alpha}{\sideset{_{#1}}{_{#2}}{\mathop{\circ}}}}
\newcommand{\oxisym}[1]{\overset{\scriptscriptstyle \omega_\alpha}{\xi_{#1}}}
\newcommand{\oAss}{\mathcal{A}\mathit{ss}}
\newcommand{\oCom}{\mathcal{C}\mathit{om}}
\newcommand{\oLie}{\mathcal{L}\mathit{ie}}
\newcommand{\oMod}{\textbf{Mod}}

\newcommand{\ket}[1]{\left|#1 \right\rangle}
\newcommand{\bra}[1]{\left \langle#1 \right|}

\newcommand{\Etree}{\begin{tikzpicture}
\path (0,0) coordinate (0);
\path (0)++(0*360/6:0.15cm) coordinate (1);
\path (0)++(1*360/6:0.15cm) coordinate (2);
\path (0)++(2*360/6:0.15cm) coordinate (3);
\path (0)++(3*360/6:0.15cm) coordinate (4);
\path (0)++(4*360/6:0.15cm) coordinate (5);
\path (0)++(5*360/6:0.15cm) coordinate (6);
\draw (0)--(2)(0)--(3)(0)--(4)(0)--(5)(0)--(6);
\path (0)++(0.3cm,0) coordinate (0');
\path (0')++(180+0*360/6:0.15cm) coordinate (1');
\path (0')++(180+1*360/6:0.15cm) coordinate (2');
\path (0')++(180+2*360/6:0.15cm) coordinate (3');
\path (0')++(180+3*360/6:0.15cm) coordinate (4');
\path (0')++(180+4*360/6:0.15cm) coordinate (5');
\path (0')++(180+5*360/6:0.15cm) coordinate (6');
\draw (0')--(2')(0')--(3')(0')--(4')(0')--(5')(0')--(6');
\path (0')++(0.05cm,0) node[anchor=west] (c) {$\scriptstyle{, \{e\} }$};
\draw (0)--(0');
\end{tikzpicture}}
\newcommand{\Eloop}{\begin{tikzpicture}
\path (0,0) coordinate (0);
\path (0)++(180+0*360/7:0.15cm) coordinate (1);
\path (0)++(180+1*360/7:0.15cm) coordinate (2);
\path (0)++(180+2*360/7:0.15cm) coordinate (3);
\path (0)++(180+2.9*360/7:0.4cm) coordinate (4);
\path (0)++(180+4.1*360/7:0.4cm) coordinate (5);
\path (0)++(180+5*360/7:0.15cm) coordinate (6);
\path (0)++(180+6*360/7:0.15cm) coordinate (7);
\path (0)++(0.15cm,0) node[anchor=west] (c) {$\scriptstyle{, \{e\} }$};
\draw (0)--(1)(0)--(2)(0)--(3)(0)--(6)(0)--(7);
\draw (0) .. controls (4) and (5) .. (0);
\end{tikzpicture}}
\newcommand{\Tree}{\begin{tikzpicture}
\path (0,0) coordinate (0);
\path (0)++(0*360/6:0.6cm) coordinate (1);
\path (0)++(1*360/6:0.6cm) coordinate (2);
\path (0)++(2*360/6:0.6cm) coordinate (3);
\path (0)++(3*360/6:0.6cm) coordinate (4);
\path (0)++(4*360/6:0.6cm) coordinate (5);
\path (0)++(5*360/6:0.6cm) coordinate (6);
\draw (0)--(2)(0)--(3)(0)--(4)(0)--(5)(0)--(6);
\path (0)++(1.1cm,0) coordinate (0');
\path (0')++(180+0*360/6:0.6cm) coordinate (1');
\path (0')++(180+1*360/6:0.6cm) coordinate (2');
\path (0')++(180+2*360/6:0.6cm) coordinate (3');
\path (0')++(180+3*360/6:0.6cm) coordinate (4');
\path (0')++(180+4*360/6:0.6cm) coordinate (5');
\path (0')++(180+5*360/6:0.6cm) coordinate (6');
\draw (0')--(2')(0')--(3')(0')--(4')(0')--(5')(0')--(6');
\draw (0)--(0');
\end{tikzpicture}}
\newcommand{\Loop}{\begin{tikzpicture}
\path (0,0) coordinate (0);
\path (0)++(180+0*360/7:0.6cm) coordinate (1);
\path (0)++(180+1*360/7:0.6cm) coordinate (2);
\path (0)++(180+2*360/7:0.6cm) coordinate (3);
\path (0)++(180+2.9*360/7:1.4cm) coordinate (4);
\path (0)++(180+4.1*360/7:1.4cm) coordinate (5);
\path (0)++(180+5*360/7:0.6cm) coordinate (6);
\path (0)++(180+6*360/7:0.6cm) coordinate (7);
\draw (0)--(1)(0)--(2)(0)--(3)(0)--(6)(0)--(7);
\draw (0) .. controls (4) and (5) .. (0);
\end{tikzpicture}}

\newtheorem{Def}{Definition}
\newtheorem{Thm}{Theorem}
\newtheorem{Lem}{Lemma}

\title{Type II Superstring Field Theory: Geometric Approach and Operadic Description}

\author[a,b]{Branislav Jur\v co}
\author[c]{Korbinian M\"unster}
\affiliation[a,1]{Mathematical Institute, Faculty of Mathematics and Physics, Charles University, Sokolovsk\'{a} 83, 186 75 Praha 8, Czech Republic}
\note{permanent address}
\affiliation[b]{CERN, Theory Division, CH-1211 Geneva 23, Switzerland}
\affiliation[c]{Arnold Sommerfeld Center for Theoretical Physics, Theresienstrasse 37, D-80333 Munich, Germany}

\emailAdd{branislav.jurco@googlemail.com}
\emailAdd{jurco@karlin.mff.cuni.cz}
\emailAdd{korbinian.muenster@physik.uni-muenchen.de}

\abstract{
We outline the construction of type II superstring field theory
leading to a geometric and algebraic BV master equation, analogous to
Zwiebach's construction for the bosonic string. The construction
uses the small Hilbert space. Elementary vertices of the non-polynomial
action are described with the help of a properly formulated minimal area
problem. They give rise to an infinite tower of superstring field
products defining a
$\mathcal{N}=1$ generalization of a loop homotopy Lie algebra, the genus zero
part generalizing a homotopy Lie algebra. Finally, we give an
operadic interpretation of the construction.
}


\arxivnumber{1303.2323}

\maketitle
\flushbottom


\section{Introduction and Summary}\label{sec:intro}
The first attempt towards a field theory of superstrings was initiated by the work of Witten \cite{Witten CSsuper}, by seeking a Chern-Simons like action for open superstrings similar to the one of open bosonic string field theory \cite{Witten CSbosonic}. The major obstacle compared to the bosonic string is the necessity of picture changing operators. Indeed, the cubic superstring theory of \cite{Witten CSsuper} turns out to be inconsistent due to singularities arising form the collision of picture changing operators \cite{Wendt}. In order to circumvent this problem, another approach was pursued which sets the string field into a different picture \cite{Thorn, Arefeva}, but upon including the Ramond sector, the modified superstring field theory suffers from similar inconsistencies \cite{Kroyter}. These two approaches are based on the small Hilbert space, the state space including the reparametrization ghosts and superghosts as they arise from gauge fixing. Upon bosonization of the superghosts, an additional zero mode arises which allows the formulation of a WZW like action for the NS sector of open superstring field theory \cite{Berkovits}. In contrast to bosonic string field theory, BV quantization of this theory is more intricate than simply relaxing the ghost number constraint for the fields of the classical action \cite{Torii, Berkovits BV}. Finally, there is a formulation of open superstring field theory that differs from all other approaches in not fixing the picture of classical fields \cite{Kroyter democratic}.

On the other hand, the construction of bosonic closed string field theory \cite{Zwiebach closed} takes its origin in the moduli space of closed Riemann surfaces. Vertices represent a subspace of the moduli space, such that the moduli space decomposes uniquely into vertices and graphs, and do not apriori require a background. Graphs are constructed from the vertices by sewing together punctures along prescribed local coordinates around the punctures. But an assignment of local coordinates around the punctures, globally on the moduli space, is possible only up to rotations. This fact implies the level matching condition and via gauge invariance also the $b_0^-=0$ constraint.

In an almost unnoticed work \cite{Yeh}, the geometric approach developed in bosonic closed string field theory, as described in the previous paragraph, has been generalized to the context of superstring field theory. Neveu-Schwarz punctures behave quite similar to punctures in the bosonic case, but a Ramond puncture describes a divisor on a super Riemann surface rather than a point. As a consequence, local coordinates around Ramond punctures, globally defined over super moduli space, can be fixed only up to rotations and translation in the Ramond divisor.

A given background provides forms on super moduli space \cite{Belopolsky geometric, Alvarez} in the sense of geometric integration theory on supermanifolds \cite{Voronov}, and in particular the geometric meaning of picture changing operators has been clarified \cite{Belopolsky picture}: Integrating along an odd direction in moduli space inevitably generates a picture changing operator. Thus, the ambiguity of defining local coordinates around Ramond punctures produces a picture changing operator associated with the vector field generating translations in the Ramond divisor. The bpz inner product plus the additional insertions originating from the sewing define the symplectic form relevant for BV quantization. As in the bosonic case, we require that the symplectic form has to be non-degenerate, but the fact that the picture changing operator present in the Ramond sector has a non-trivial kernel, forces to impose additional restrictions besides the level matching and $b_0^-=0$ constraint on the state space.

The purpose of this paper is to describe the construction of type II superstring field theory in the geometric approach. We start in section \ref{sec:geobv} by defining a BV structure on the moduli space of type II world sheets decorated with coordinate curves. A coordinate curve determines local coordinates around the punctures up to rotations and translations in the Ramond divisors. The BV operator and the antibracket correspond to the sewing of punctures along coordinate curves in the non-separating (both punctures on a single connected world sheet) and separating (punctures located on two disconnected world sheets) case respectively.

In section \ref{sec:algbv}, we then review the operator formalism in the context of superstrings and the construction of forms on super moduli space. We define the symplectic form in the various sectors and determine the corresponding restricted state spaces. The symplectic form induces a BV structure on the space of multilinear maps on the restricted state spaces, and the factorization and chain map properties of the forms make the combined superconformal field theory of the matter and ghost sector a morphism of BV algebras. Note that the relevant grading in the BV formalism is the ghost number but not the picture.

Finally, we propose a minimal area problem in section \ref{sec:vertices}, which determines the geometric vertices of type II superstring field theory and furthermore induces a section from the super moduli space to the super moduli space decorated with coordinate curves. The requirement that Feynman graphs produce a single cover of moduli space implies that the geometric vertices satisfy the BV master equation. For a given background, the algebraic vertices are defined by integrating the geometric vertices w.r.t. the corresponding forms, and satisfy the BV master equation as well. The kinetic term of the theory is given by the symplectic form together with the BRST charge.

The construction of string field theory in the geometric approach manifestly leads to a BV master equation on the moduli space, which describes the background independent part of string field theory. The second ingredient is a background, which defines a morphism of BV algebras. In section \ref{sec:operad}, we elucidate the relevance of operads in the context of string field theory. The usefulness of operads in formulating string field theory derives from a theorem due to Barannikov \cite{Barannikov operadBV}, which establishes a one-to-one correspondence between morphisms over the Feynman transform of a modular operad and solutions to an associated BV master equation. We conclude that the decomposition of the moduli space into vertices and graphs defines a morphism from the Feynman transform of the modular operad encoding the symmetry properties of the vertices to the chain complex of moduli spaces. A background then corresponds to a morphism from the chain complex of moduli spaces to the endomorphism operad whose vector space is the state space, the differential is the BRST charge and the contraction maps are defined w.r.t. the symplectic form. Altogether, the composition of these two morphisms determines the algebraic structure of the vertices. In closed string field theory the vertices satisfy the axioms of a loop homotopy Lie-algebra \cite{Markl loop}, whose tree-level part is a homotopy Lie-algebra ($L_\infty$-algebra). We introduce the relevant operad for type II superstring field theory and define algebras over its Feynman transform to be $\mathcal{N}=1$ loop homotopy Lie-algebras.

Appendix \ref{app:srs} includes a brief account of super Riemann surfaces, in order to make the paper self contained. In appendix \ref{app:scft}, we treat the superconformal field theory of type II superstring theory, with a particular focus on the ghost sector. We define ghost number and picture in an unconventional way, avoiding half integer picture number in the Ramond sector. Finally, appendix \ref{app:forms}, reviews the geometric integration theory on supermanifolds and its relation to superstring theory, following \cite{Belopolsky picture,Belopolsky geometric}.

\section{Supermoduli Space and Geometric BV Structure}\label{sec:geobv}
The basic requirement of string field theory is, that its vertices reproduce the perturbative string amplitudes via Feynman rules. The fundamental object of interest is thus the appropriate moduli space of world sheets. Following \cite{Witten srs, Witten integration}, a type II world sheet $\bf{\Sigma}$ is a smooth supermanifold embedded in $\Sigma \times \ti{\Sigma}$, where $\Sigma$ and $\ti{\Sigma}$ are super Riemann surfaces s.t. the reduced space of $\ti{\Sigma}$ is the complex conjugate of the reduced space of $\Sigma$. We refer to $\Sigma$ as the holomorphic and $\ti{\Sigma}$ as the antiholomorphic sector, in analogy to the bosonic case. We require that the total number of punctures on $\Sigma$ and $\ti{\Sigma}$ coincide, but not that the number of punctures for NS and R coincide separately. Furthermore there is no condition imposed on the spin structures. The dimension of $\bf{\Sigma}$ is $2|2$, whereas the dimension of $\Sigma \times \ti{\Sigma}$ as a smooth supermanifold is $4|2$. Conversely, given reduced spaces $\Sigma_{\op{red}}$ and $\ti{\Sigma}_{\op{red}}$ which are complex conjugate to each other, $\bf{\Sigma}$ can be constructed by thickening the diagonal of $\Sigma_{\op{red}}\times  \ti{\Sigma}_{\op{red}}$ in the odd directions. The operation of thickening in the odd directions is unique up to homology, which is good enough since the world sheet action is defined by integrating $\bf{\Sigma}$ over a closed form.

The moduli space of super Riemann surfaces of genus $g$ with $n_{NS}$ NS punctures and $n_R$ Ramond punctures is denoted by $\mathfrak{M}_{g,n_R,n_{NS}}$. Its complex dimension is
\be\no
\op{dim}(\mathfrak{M}_{g,n_R,n_{NS}})= 3g-3+n_{NS}+n_R\;|\;2g-2+n_{NS}+\tinv{2}n_{R} \pt
\ee
This is not quite the appropriate moduli space for type II strings. We need a moduli space that parametrizes inequivalent type II world sheets, and thus we proceed as in the previous paragraph:
Consider the reduced space $\bracketi{\mathfrak{M}_{g,\nn,\nr}}_{\op{red}}$ and its complex conjugate $\bracketi{\ti{\mathfrak{M}}_{g,{\tnn},{\tnr}}}_{\op{red}}$. The moduli space of type II strings $\mathfrak{M}^{II}_{g,\vect{n}}$ is defined by thickening the diagonal of $\bracketi{\mathfrak{M}_{g,\nn,\nr}}_{\op{red}}\times \bracketi{\ti{\mathfrak{M}}_{g,{\tnn},{\tnr}}}_{\op{red}}$ in the odd directions. Again this operation is unique up to homology, but since superstring amplitudes are defined by integrating $\mathfrak{M}^{II}_{g,\vect{n}}$ over a closed form, this ambiguity does not matter.
We have four different kinds of punctures
\be\no
\vect{n}=(\nnn,\nnr,\nrn,\nrr)
\ee
satisfying
\begin{align}\label{eq:cond}
\nnn+\nnr&=\nn \in \N_0\\\no
\nnn+\nrn&=\tnn  \in \N_0\\\no
\nrr+\nrn&=\nr \in 2\N_0 \\\no
\nrr+\nnr&=\tnr  \in 2\N_0 \pt
\end{align}
Thus we conclude that the dimension of $\mathfrak{M}^{II}_{g,\vect{n}}$ as a smooth supermanifold is given by
\be\no
\op{dim}(\mathfrak{M}^{II}_{g,\vect{n}})= 6g-6+2n \;|\; 4g-4 +2\nnn +\tfrac{3}{2}(\nnr+\nrn)+\nrr \pt
\ee

This describes the geometric data which is needed to define superstring perturbation theory. In a field theory formulation of string theory, however, we need additional structure. Vertices represent a subspace of the full moduli space, and Feynman graphs are constructed by sewing surfaces along punctures. To perform the sewing operation, we have to know which points in a neighborhood of one puncture to identify with which points in a neighborhood of the other puncture. The required extra structure is that of a \emph{coordinate curve} around each puncture, which is an embedded submanifold $S^{1|2}_\alpha \subset \Sigma$ encircling a single puncture of type $\alpha\in \{{ NS-NS,\;NS-R,\;R-NS,\;R-R}\}$, where $S^{1|2}_\alpha$ is the supercircle with two odd directions and boundary condition $\alpha$. Such a coordinate curve determines a local superconformal coordinate system $(z,\tilde{z},\vt,\tilde{\vt})$, where the puncture is located at
\begin{align}\no
(z,\tilde{z},\vt,\tilde{\vt})=0 &\sep NS-NS \\\no
(z,\tilde{z},\tilde{\vt})=0 &\sep R-NS \\\no
(z,\tilde{z},\vt)=0 &\sep NS-R \\\no
(z,\tilde{z})=0 &\sep R-R \com
\end{align}
up to rotations generated by $l_0^{-}\defineL l_0-\ti{l}_0$ and translations in the Ramond divisors (if present) generated by $g_0$ and $\ti{g}_0$.

We denote the moduli space of type II world sheets decorated with \emph{coordinate curves} by $\hat{\mathfrak{P}}^{II}_{g,\vect{n}}\,$, whereas the moduli space decorated with \emph{local coordinates} is denoted by ${\mathfrak{P}}^{II}_{g,\vect{n}}\,$.  The decorated spaces are of course infinite dimensional and can be considered as a fibre bundle over $\mathfrak{M}^{II}_{g,\vect{n}}$ by discarding the information about the coordinate curves/local coordinates. In section \ref{sec:vertices}, we propose that $\hat{\mathfrak{P}}^{II}_{g,\vect{n}}$ is indeed a trivial bundle, by outlining the construction of a global section. In contrast, the moduli space ${\mathfrak{P}}^{II}_{g,\vect{n}}\,$ does not admit global sections \cite{Zwiebach closed}.

We will start by defining the sewing operations for given local coordinate systems: Consider two punctures $p$ and $p^\prime$ of the same type, together with local coordinates $(z,\tilde{z},\vt,\tilde{\vt})$ and $(z^\prime,\tilde{z}^\prime,\vt^\prime,\tilde{\vt}^\prime)$. The punctures may either reside on a single connected surface or on two disconnected surfaces, which we call the non-separating and separating case respectively. First we will focus on the holomorphic sector. In the bosonic case, the sewing operation for two given coordinate systems $z$ and $z^\prime$ is given by the identification
\be\label{eq:bpzbosonic}
z^\prime=I(z)\defineL-\frac{1}{z} \pt
\ee

\begin{figure}[h] 
\begin{center}
\begin{minipage}[b]{0.4\textwidth}
\resizebox{6cm}{!}{
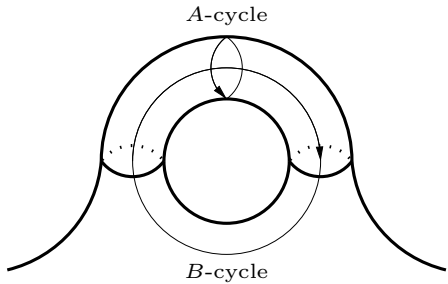}
\end{minipage}
\end{center}
\caption{Sewing operation in the non-separating case.}
\label{fig:nonsep}
\end{figure}

From equation (\ref{eq:mobns}), we can infer that the generalization of the sewing map (\ref{eq:bpzbosonic}) for the NS sector is given by
\be\label{eq:bpznspm}
I_{(\pm,+)}(z,\theta)=\bpm
\displaystyle -\frac{1}{z} \vspace{0.1cm} \\ \vspace{0.1cm}\displaystyle \pm\frac{\vt}{z}
\epm \pt
\ee
In the separating case, there is no essential difference between $I_{(+,+)}$ and $I_{(-,+)}$, they are related by replacing $\vt\to-\vt$ on one surface globally. For the non-separating case the situation is different. Assume that a transition from $(z,\vt)$ to $(z^\prime,\vt^\prime)$ does not change the sign in the odd coordinate, i.e. that for a coordinate system $(z^{\prime\prime},\vt^{\prime\prime})$ covering $(z,\vt)$ and $(z^\prime,\vt^\prime)$, the transition functions from $(z^{\prime\prime},\vt^{\prime\prime})$ to $(z,\vt)$ and $(z^{\prime\prime},\vt^{\prime\prime})$ to $(z^\prime,\vt^\prime)$ are both of the form (\ref{eq:confns}) with the same sign in front of $\vt^{\prime\prime}$. Under this assumption, the sewing with $I_{(\pm,+)}$ generates a handle with $\pm$ spin structure along the $B$-cycle, see figure \ref{fig:nonsep}.

In the R sector, the sewing map follows from generalizing (\ref{eq:bpzbosonic}) according to (\ref{eq:mobr}):
\be\label{eq:bpzrpm}
I_{(\pm,-)}(z,\theta)=\bpm
\displaystyle -\frac{1}{z} \vspace{0.1cm} \\ \vspace{0.1cm} \displaystyle\pm i \vt
\epm \pt
\ee
Similarly as in the NS sector, the sewing with $I_{(\pm,-)}$ in the non-separating case generates a handle with $\pm$ spin structure along the $B$-cycle. For the $A$-cycle, the $+$ and $-$ spin structure corresponds to NS and R respectively, which justifies the notation.

Modular invariance requires a sum over all spin structures. The modular invariant combination of spin structures is known to be
\be\no
(+,+)-(-,+)-(+,-)\pm(--) \pt
\ee
Thus we can determine the sewing operations to be
\be\label{eq:bpzns}
I_{NS}=\inv{2}\bracket{I_{(+,+)}-I_{(-,+)}}= \Pi^{GSO^-}\circ I_{(+,+)}
\ee
and
\be\label{eq:bpzr}
I_{R}^\pm=\inv{2}\bracket{I_{(+,-)}\pm I_{(-,-)}}= \Pi^{GSO^{\pm}}\circ I_{(+,-)}
\ee
for the NS sand R sector, respectively. In equation (\ref{eq:bpzns}) and (\ref{eq:bpzr}), the sum has to be understood as generating two surfaces from a given one and taking their formal linear combination, which defines the GSO projection $\Pi^{GSO^\pm}$. These are the maps that determine the bpz conjugation in superconformal field theory (see appendix \ref{app:scft}).
Combining the holomorphic and antiholomorphic sector, we end up with
\be\label{eq:bpzgeo}
I_\alpha(z,\ti{z},\vt,\ti{\vt}) \sep \alpha\in \{{ NS-NS,\;NS-R,\;R-NS,\;R-R}\} \pt
\ee

Now let us describe the sewing operation for given coordinate curves. As discussed previously, a coordinate curve does not uniquely determine a local coordinate system. This ambiguity naturally leads to a family of surfaces associated to the sewing of two punctures. We begin by restricting our considerations to the holomorphic sector. In the NS sector the local coordinate system is determined up to rotations generated by $l_0-\ti{l}_0$. Let $\varphi^{l_0}_t$ be the flow generated by $l_0$,
\be\no
\pd_t\varphi^{l_0}_t=l_0\circ \varphi^{l_0}_t \com
\ee
which leads to
\be\no
\varphi^{l_0}_t(z,\vt)= \bpm
e^{-t}z\\
e^{-t/2}\vt
\epm
\pt
\ee
The family of local coordinate systems associated to a coordinate curve in the NS sector is parametrized by an angle $\vartheta\in[0,2\pi]$ and the corresponding sewing operation is given by
\be\label{eq:sewingns}
\phi_\vartheta= I_{NS}\circ \varphi^{l_0}_{i\vartheta}= \Pi^{GSO^-}\circ I_{(+,+)}\circ \varphi^{l_0}_{i\vartheta} \com
\ee
which explicitly reads
\be\label{eq:sewingnsex}
I_{(+,+)}\circ \varphi^{l_0}_{i\vartheta}(z,\vt)=\bpm
\displaystyle -\frac{e^{i\vartheta}}{z} \vspace{0.1cm} \\ \vspace{0.1cm}
\displaystyle \vt \frac{e^{i\vartheta/2}}{z}
\epm
\pt
\ee
In the R sector the local coordinate system is determined up to rotations and translations in the Ramond divisor generated by $g_0$. Let $\varphi^{g_0}_{t,\tau}$ be the flow generated by $g_0$,
\be\no
(\pd_\tau +\tau\pd_t)\varphi^{g_0}_{t,\tau}=g_0\circ \varphi^{g_0}_{t,\tau}
\ee
which leads to
\be\no
\varphi^{g_0}_{t,\tau}(z,\vt)= \bpm
e^{-t}z(1+\vt \tau)\\
\vt + \tau
\epm
\pt
\ee
We conclude, that in the R sector the family of local coordinate systems associated to a coordinate curve is parametrized by an angle $\vartheta\in[0,2\pi]$ and an odd parameter $\tau\in \C^{0|1}$, and the corresponding sewing operation reads
\be\label{eq:sewingr}
\phi_{\vartheta,\tau}^\pm= I_{R}^\pm\circ \varphi^{g_0}_{i\vartheta,\tau}= \Pi^{GSO^\pm}\circ I_{(+,-)}\circ \varphi^{g_0}_{i\vartheta,\tau} \pt
\ee
Explicitly, we have
\be\label{eq:sewingrex}
 I_{(+,-)}\circ \varphi^{g_0}_{i\vartheta,\tau}(z,\vt)=\bpm
\displaystyle-\frac{e^{i\vartheta}}{z}(1-\vt \tau) \vspace{0.1cm} \\ \vspace{0.1cm}
\displaystyle i(\vt+\tau)
\epm
\pt
\ee
Combining holomorphic and antiholomorphic sectors, we identify the four sewing operations to be
\begin{align}\label{eq:sewing}
(\Phi_{NS-NS})_\vartheta&=\bracketi{ I_{NS}\circ \varphi^{l_0}_{i\vartheta} \,,\, \ti{I}_{NS}\circ \varphi^{\ti{l}_0}_{-i\vartheta}} \\\no
(\Phi_{R-NS})_{\vartheta,\tau}&=\bracketi{I_{R}\circ \varphi^{g_0}_{i\vartheta,\tau} \,,\, \ti{I}_{NS}\circ \varphi^{\ti{l}_0}_{-i\vartheta}}  \\\no
(\Phi_{NS-R})_{\vartheta,\ti{\tau}}&=\bracketi{I_{NS}\circ \varphi^{{l}_0}_{i\vartheta}  \,,\, \ti{I}_{R}\circ \varphi^{\ti{g}_0}_{-i\vartheta,\ti{\tau}}}  \\\no
(\Phi_{R-R})_{\vartheta,\tau,\ti{\tau}}&=\bracketi{I_{R}\circ \varphi^{{g}_0}_{i\vartheta,\tau}  \,,\, \ti{I}_{R}\circ \varphi^{\ti{g}_0}_{-i\vartheta,\ti{\tau}}} \pt
\end{align}

The geometric vertices of string field theory represent a subspace of the full moduli space. Thus the natural object to consider is the singular chain complex
\be\label{eq:cc}
C^{\bullet|\bullet}(\hat{\mathfrak{P}}^{II}_{g,\vect{n}})\pt
\ee
The grading for $\mathcal{A}_{g,\vect{n}}\in C^{k|l}(\hat{\mathfrak{P}}^{II}_{g,\vect{n}})$ is defined by codimension, i.e.
\be\label{eq:grad}
k|l=\op{deg}(\mathcal{A}_{g,\vect{n}})\defineL \op{dim}(\mathfrak{M}^{II}_{g,\vect{n}})-\op{dim}(\mathcal{A}_{g,\vect{n}}) \pt
\ee
Furthermore we endow the chains with an orientation. In the context of supergeometry, there are different notions of orientation on a supermanifold $M^{m|n}$, corresponding to the four normal subgroups of the general linear group $\op{GL}(m|n)$, described in appendix \ref{app:forms}. The relevant notion for integrating forms is that of a $[+-]$ orientation, see e.g. \cite{Voronov} or appendix \ref{app:forms}, which requires $\op{det}(g_{00})>0$ for
\be\no
\op{GL}(m|n)\ni g=\bpm g_{00} & g_{01}\\ g_{10} & g_{11} \epm \pt
\ee

Now we are going to describe the BV structure on the chain complex of moduli spaces. The final aim is of course to dress the punctures with vertex operators, which forces us to implement the indistinguishability of identical particles already at the geometric level. We proceed as follows: We define
\be\no
\oMod(\oCom^{\mathcal{N}=1})(g,\vect{n})
\ee
to be a one dimensional vector space\footnote{The notation for this object will be justified in section \ref{sec:operad}, where we introduce operads and explain their applications to string field theory.}.
Furthermore, the permutation group $\Sigma_{\vect{n}}\defineL \times_\alpha \Sigma_{n_\alpha}$ acts on $\oMod(\oCom^{\mathcal{N}=1})(g,\vect{n})$ by the trivial representation. According to the geometrical interpretation, we require $g\ge0$ and the conditions of (\ref{eq:cond}). Hence, the chains with appropriate symmetry properties can be described by the invariants
\be\label{eq:icc}
C^{\bullet|\bullet}_{\op{inv}}(\hat{\mathfrak{P}}^{II}_{g,\vect{n}}) \defineL\bracketii{C^{\bullet|\bullet}(\hat{\mathfrak{P}}^{II}_{g,\vect{n}})\tp\oMod(\oCom^{\mathcal{N}=1})(g,\vect{n})}^{\Sigma_{\vect{n}}} \com
\ee
where the permutation group $\Sigma_{\vect{n}}$ acts on $C^{\bullet|\bullet}(\hat{\mathfrak{P}}^{II}_{g,\vect{n}})$ by permutation of punctures. We call (\ref{eq:icc}) the invariant chain complex.
All that is just saying, that we restrict to chains which are invariant under permutations of punctures of the same type.

Let $\oophi{i}{j}$ be the sewing operation in the separating case. The input of $\oophi{i}{j}$ is a pair of surfaces decorated with coordinate curves, and its output is the family of surfaces generated by sewing together puncture $i$ on the first surface with puncture $j$ on the second surface according to (\ref{eq:sewing}), where both punctures $i,j$ are of type $\alpha$.
Analogously, we define $\oxiphi{ij}$ to be the sewing operation in the non-separating case. For later use, we furthermore define maps $\ooI{i}{j}$ and $\oxiI{ij}$, involving the sewing (\ref{eq:bpzgeo}) suitable for surfaces decorated with local coordinate around the punctures. The two former operations induce maps on the chain complex (\ref{eq:cc}), which we also denote by $\oophi{i}{j}$ and $\oxiphi{ij}$, by defining their action pointwise. From (\ref{eq:sewing}) and the definition of the grading (\ref{eq:grad}), we conclude that for all $\alpha$, $\oophi{i}{j}$ and $\oxiphi{ij}$ are of degree $1|0$, that is
\be\label{eq:sewingtree}
\oophi{i}{j}:  C^{k_1|l_1}(\hat{\mathfrak{P}}^{II}_{g_1,\vect{n}_1+e_\alpha}) \times C^{k_2|l_2}(\hat{\mathfrak{P}}^{II}_{g_2,\vect{n}_2+e_\alpha}) \to C^{k_1+k_2+1|l_1+l_2}(\hat{\mathfrak{P}}^{II}_{g_1+g_2,\vect{n}_1+\vect{n}_2})
\ee
and
\be\label{eq:sewingloop}
\oxiphi{ij}:  C^{k|l}(\hat{\mathfrak{P}}^{II}_{g,\vect{n}+2e_\alpha}) \to C^{k+1|l}(\hat{\mathfrak{P}}^{II}_{g+1,\vect{n}})\com
\ee
where $e_\alpha$ denotes the unit vector in direction $\alpha$ and represents puncture $i$ respectively $j$. Note also that the boundary operator
\be\no
\pd: C^{k|l}(\hat{\mathfrak{P}}^{II}_{g,\vect{n}})\to C^{k+1|l}(\hat{\mathfrak{P}}^{II}_{g,\vect{n}})
\ee
is of degree $1|0$ due to the choice of grading.

Finally, we want to lift $\oophi{i}{j}$ and $\oxiphi{ij}$ to maps on the invariant chain complex (\ref{eq:icc}), which will lead to the desired BV structure.
Let $\mathcal{B}_{g_1,\vect{n}_1+e_\alpha}\in C^{k_1|l_1}_{\op{inv}}(\hat{\mathfrak{P}}^{II}_{g_1,\vect{n}_1+e_\alpha})$ and $\mathcal{B}_{g_2,\vect{n}_2+e_\alpha} \in C^{k_2|l_2}_{\op{inv}}(\hat{\mathfrak{P}}^{II}_{g_2,\vect{n}_2+e_\alpha})$ be invariant chains and consider the expression
\begin{align}\label{eq:geoop}
(\mathcal{B}_{g_1,\vect{n}_1+e_\alpha},\mathcal{B}_{g_2,\vect{n}_2+e_\alpha})^{\op{geo}}_\alpha
&\defineL \sum_{\sigma\in\op{sh}\bracket{\vect{n}_1,\vect{n}_2}} \sigma.\bracketi{\mathcal{B}_{g_1,\vect{n}_1+e_\alpha} \oophi{i}{j} \mathcal{B}_{g_2,\vect{n}_2+e_\alpha}}  \com\\\no
(\cdot,\cdot)^{\op{geo}} &\defineL \sum_\alpha (\cdot,\cdot)^{\op{geo}}_\alpha \pt
\end{align}
First, note that since $\mathcal{B}_{g_1,\vect{n}_1+e_\alpha}$ and $\mathcal{B}_{g_2,\vect{n}_2+e_\alpha}$ are invariant under permutation of punctures of the same type, it does not matter which punctures $i$ and $j$ we choose for the sewing operation $\oophi{i}{j} $. That is why $i$ and $j$ does not appear on the left hand side of (\ref{eq:geoop}). Second, $\op{sh}\bracket{\vect{n}_1,\vect{n}_2}$ denotes the set of shuffles\footnote{The set of shuffles $\op{sh}(n,m)\subset\Sigma_{n+m}$ contains all permutation $\sigma\in\Sigma_{n+m}$, satisfying $\sigma_1<\dots<\sigma_n$ and $\sigma_{n+1}<\dots<\sigma_{n+m}$.} of the punctures $\vect{n}_1$ and $\vect{n}_2$ that remain after sewing.
In the non-separating case, we define
\begin{align}\label{eq:geobv}
\Delta^{\op{geo}}_\alpha \mathcal{B}_{g,\vect{n}+2e_\alpha}  &\defineL \oxiphi{ij} \bracket{\mathcal{B}_{g,\vect{n}+2e_\alpha}}    \com\\\no
\Delta^{\op{geo}}  &\defineL \sum_\alpha\Delta^{\op{geo}}_\alpha \pt
\end{align}
for $\mathcal{B}_{g,\vect{n}+2e_\alpha}\in C^{k|l}_{\op{inv}}(\hat{\mathfrak{P}}^{II}_{g,\vect{n}+2e_\alpha})$.
Again the $\Sigma_{\vect{n}}$ invariance guarantees independence of the choice of punctures $i$ and $j$.

Now one can show that $\pd$, $(\cdot,\cdot)^{\op{geo}}$ and $\Delta^{\op{geo}}$ satisfy the axioms of a differential BV algebra, that is (leaving out the superscript $\op{geo}$)
\begin{align}\label{eq:axiomsbv}
\pd^2&=0\\\no
\Delta^2&=0\\\no
\pd\Delta+\Delta\pd&=0\\\no
\pd\circ (\cdot,\cdot)&=(\pd,\cdot)-(\cdot,\pd)\\\no
\Delta\circ (\cdot,\cdot)&=(\Delta,\cdot)-(\cdot,\Delta)\\\no
(\mathcal{B}_{g_1,\vect{n}_1},\mathcal{B}_{g_2,\vect{n}_2})&= -(-1)^{(k_1+1)(k_2+1)}(\mathcal{B}_{g_2,\vect{n}_2},\mathcal{B}_{g_1,\vect{n}_1})\\\no
(-1)^{(k_1+1)(k_3+1)}&((\mathcal{B}_{g_1,\vect{n}_1},\mathcal{B}_{g_2,\vect{n}_2}),\mathcal{B}_{g_3,\vect{n}_3}))+\op{cycl.}=0 \com
\end{align}
where $\mathcal{B}_{g_i,\vect{n}_i}\in C^{k_i|l_i}_{\op{inv}}(\hat{\mathfrak{P}}^{II}_{g_i,\vect{n}_i})$. Note that only the even part $k$ of the grading $k|l$ enters in the expressions for the signs, thus the odd part $l$ is not really a grading in the strict sense, it is merely an additional index representing the odd codimensionality of the chain. The reason for this resides in the fact that we chose the $[+-]$ orientation for the chains.
The proof of the identities (\ref{eq:axiomsbv}) follows directly from the proof in the bosonic case \cite{Zwiebach closed, Zwiebach background}, again due to the choice of orientation: The $[+-]$ orientation distinguishes an order for the even vectors but not for the odd vectors. For all $\alpha$, the operations $\oophi{i}{j}$ and $\oxiphi{ij}$ increase the even dimensionality by one due to the twist angle $\vartheta$ and thus the proof of (\ref{eq:axiomsbv}) reduces to that in the bosonic case.

Indeed, a BV algebra also requires a graded commutative multiplication, such that $\Delta$ defines a second order derivation and $\pd$ a first order derivation. We do not describe this operation here, but definitely it can be defined similarly to the bosonic case by disjoint union \cite{Zwiebach closed, Zwiebach background}.

\section{Operator Formalism and Algebraic BV structure}\label{sec:algbv}
The geometric BV algebra discussed in the previous section describes the background independent ingredient of type II superstring field theory. A background refers to a superconformal field theory (SCFT) with additional structure provided by the superconformal ghosts and the BRST charge, which allows the construction of a measure on supermoduli space compatible with the sewing operations. Such a field theory is called a topological superconformal field theory (TSCFT) \cite{Belopolsky geometric, Getzler bv}.

We start by introducing differential forms on supermoduli space, following \cite{Alvarez, Belopolsky geometric}. Let $\mathcal{H}_\alpha$, $\alpha\in\{{ NS-NS,\;NS-R,\;R-NS,\;R-R}\}$, denote the state spaces of a type II SCFT (see appendix \ref{app:scft}).  For a given type II world sheet $\B{\Sigma}_{g,\vect{n}}\in {\mathfrak{P}}^{II}_{g,\vect{n}}$ with local coordinates around the punctures, the SCFT assigns a multilinear map
\be\no
Z(\B{\Sigma}_{g,\vect{n}}):\mathcal{H}^{\tp \vect{n}}\to \C^{1|1} \com
\ee
where
\be\no
\mathcal{H}^{\tp \vect{n}}\defineL \bigotimes_\alpha  (\mathcal{H}_\alpha)^{\tp n_\alpha} \pt
\ee
Let $\oobpz{i}{j}$ be the map
\be\no
\oobpz{i}{j}: \op{Hom}(\mathcal{H}^{\tp \vect{n}_1+e_\alpha},\C^{1|1}) \times \op{Hom}(\mathcal{H}^{\tp \vect{n}_2+e_\alpha},\C^{1|1}) \to \op{Hom}(\mathcal{H}^{\tp \vect{n}_1+\vect{n}_2},\C^{1|1})
\ee
 that contracts input $i$ of the first linear map with input $j$ of the second linear map, both of type $\alpha$, w.r.t. the inverse of the bpz inner product $\op{bpz}_\alpha^{-1}$.
Analogously, we define the map
 \be\no
 \oxibpz{ij}:\op{Hom}(\mathcal{H}^{\tp \vect{n}+2e_\alpha},\C^{1|1}) \to \op{Hom}(\mathcal{H}^{\tp \vect{n}},\C^{1|1}) \pt
 \ee
 The factorization properties
\be\label{eq:factorizationbpz1}
Z\bracketi{\B{\Sigma}_{g_1,\vect{n}_1+e_\alpha}\,\ooI{i}{j}\, \B{\Sigma}_{g_2,\vect{n}_2+e_\alpha}}=Z\bracketi{\B{\Sigma}_{g_1,\vect{n}_1+e_\alpha}}\,\oobpz{i}{j}\, Z\bracketi{\B{\Sigma}_{g_2,\vect{n}_2+e_\alpha}}
\ee
and
\be\label{eq:factorizationbpz2}
Z\bracketi{\oxiI{ij}\,\B{\Sigma}_{g_1,\vect{n}+2e_\alpha}}=\oxibpz{ij}\,Z\bracketi{\B{\Sigma}_{g_1,\vect{n}+2e_\alpha}}
\ee
hold, with the sewing operations $\ooI{i}{j}$ and $\oxiI{ij}$ introduced in the previous section. Furthermore the tensor structure is preserved, i.e.
\be\no
Z(\B{\Sigma}_{g_1,\vect{n}_1}\sqcup \B{\Sigma}_{g_2,\vect{n}_2})=Z(\B{\Sigma}_{g_1,\vect{n}_1})\tp Z(\B{\Sigma}_{g_2,\vect{n}_2}) \pt
\ee

A tangent vector $V\in T{\mathfrak{P}}^{II}_{g,\vect{n}}$ can be represented by a collection of pairs of holomorphic and antiholomorphic Virasoro vectors $\vect{v}=\bracketi{(v^{(1)},\ti{v}^{(1)}),\dots,(v^{(n)},\ti{v}^{(n)})}$, $n=\sum_\alpha n_\alpha$, via Schiffer variation. We can think of $Z$ as a function on ${\mathfrak{P}}^{II}_{g,\vect{n}}$ with values in $\op{Hom}(\mathcal{H}^{\tp \vect{n}},\C^{1|1})$. The relation between the tangent vector $V$ and its representation via Virasoro vectors is expressed by the relation
\be\label{eq:schiffer}
V(Z)= Z\circ T(\vect{v}) \com
\ee
where
\be\no
T(\vect{v})\defineL \sum_{i=1}^n \bracketi{T^{(i)}(v^{(i)})+ \ti{T}^{(i)}(\ti{v}^{(i)})}\com
\ee
and
\begin{align}\no
T(l_n)&=L_n\com\\\no
T(g_n)&=G_n
\end{align}
defines $T(v)$ by linearity.
Similarly, $B(v)$ is determined by
\begin{align}\no
B(l_n)&=b_n \com \\\no
B(g_n)&=\beta_n \pt
\end{align}
Furthermore, $Z$ is BRST closed and a map of Lie algebras, that is
\begin{align}\no
[V_1,V_2](Z)&=Z\circ T([\vect{v}_1,\vect{v}_2]) \com\\\no
Z\circ\sum_{i=1}^n Q^{(i)}&=0 \pt
\end{align}

Utilizing the $B$ ghost, we can now define differential forms $\omega^{k|l}_{g,\vect{n}}$ on ${\mathfrak{P}}^{II}_{g,\vect{n}}$ with values in $\op{Hom}(\mathcal{H}^{\tp \vect{n}},\C^{1|1})$ \cite{Belopolsky geometric}:
Let $(V_1,\dots,V_r|\mathcal{V}_1,\dots,\mathcal{V}_s)$, be a collection of $r$ even and $s$ odd tangent vector to ${\mathfrak{P}}^{II}_{g,\vect{n}}$ at $\B{\Sigma}_{g,\vect{n}}\,$, we define
\be\label{eq:forms}
\omega^{k|l}_{g,\vect{n}}(V_1,\dots,V_r|\mathcal{V}_1,\dots,\mathcal{V}_s)\defineL N_{g,\vect{n}}\cdot Z(\B{\Sigma}_{g,\vect{n}})\circ B(\vect{v}_1)\dots B(\vect{v}_r) \, \delta(B(\vect{\nu}_1))\dots \delta(B(\vect{\nu}_s)) \com
\ee
where $r|s=\op{dim}({\mathfrak{M}}^{II}_{g,\vect{n}})-k|l$, in accordance with the grading (\ref{eq:grad}) introduced for the chain complex of moduli spaces. The normalization constant $N_{g,\vect{n}}=(2\pi i)^{-(3g-3+n)}$ derives from the twist angle $\vartheta$ of the sewing operations (\ref{eq:sewing}) \cite{Zwiebach closed}. From (\ref{eq:degZ}) and (\ref{eq:forms}), we conclude that $\omega^{k|l}_{g,\vect{n}}$ has ghost number and picture equal to
\be\label{eq:degreeforms}
k-2n|l-2n_{NS-NS}- (n_{NS-R}+n_{R-NS}) \pt
\ee
Moreover, the differential forms define chain maps in the sense that
\be\label{eq:chainmapproperty}
d\omega^{k+1|l}_{g,\vect{n}}=(-1)^k\omega^{k|l}_{g,\vect{n}}\circ \sum_{i=1}^n Q^{(i)} \pt
\ee

Indeed, we would like to be able to pull this structure back to the finite dimensional moduli space $\mathfrak{M}^{II}_{g,\vect{n}}\,$. That is we need a natural way to assign local coordinates to type II world sheets, or in other words we require a global section of $\mathfrak{P}^{II}_{g,\vect{n}}$ as a fibre bundle over $\mathfrak{M}^{II}_{g,\vect{n}}\,$. As indicated in section \ref{sec:geobv}, the topology of $\mathfrak{P}^{II}_{g,\vect{n}}$ does not admit global sections. The best we can get are global sections of $\hat{\mathfrak{P}}^{II}_{g,\vect{n}}\,$. In section \ref{sec:vertices}, we outline the construction of a global section $\sigma$ of $\hat{\mathfrak{P}}^{II}_{g,\vect{n}}$ as a fibre bundle over ${\mathfrak{M}}^{II}_{g,\vect{n}}$ via an analog of minimal area metrics in the superconformal setting.
Therefore, in order to make use of the section $\sigma$, we first have to explain how to employ the forms (\ref{eq:forms}) in the context of $\hat{\mathfrak{P}}^{II}_{g,\vect{n}}\,$. It turns out that using $\hat{\mathfrak{P}}^{II}_{g,\vect{n}}$ instead of ${\mathfrak{P}}^{II}_{g,\vect{n}}$ requires to restrict the state spaces $\mathcal{H}_\alpha$ in a certain way \cite{Zwiebach closed, Yeh}, which we will denote by $\hat{\mathcal{H}}_\alpha$. The constraints leading to $\hat{\mathcal{H}}_\alpha$ follow from requiring factorization properties analogously to (\ref{eq:factorizationbpz1}) and (\ref{eq:factorizationbpz2}) \cite{Yeh}:
\be\label{eq:factorizationsym1}
\int\limits_{\mathcal{A}_{g_1,\vect{n}_1+e_\alpha} \,\oophi{i}{j}\, \mathcal{A}_{g_1,\vect{n}_1+e_\alpha}} \omega^{k_1+k_2+1|l_1+l_2}_{g_1+g_2,\vect{n}_1+\vect{n}_2}
=\bracketiii{\int\limits_{\mathcal{A}_{g_1,\vect{n}_1+e_\alpha}} \omega^{k_1|l_1}_{g_1,\vect{n}_1+e_\alpha}}
\,\oosym{i}{j}\,
\bracketiii{\int\limits_{\mathcal{A}_{g_2,\vect{n}_2+e_\alpha}} \omega^{k_2|l_2}_{g_2,\vect{n}_2+e_\alpha}} \com
\ee
and
\be\label{eq:factorizationsym2}
\int\limits_{\oxiphi{ij}\,\mathcal{A}_{g-1,\vect{n}+2e_\alpha}} \omega^{k+1|l}_{g,\vect{n}} = \oxisym{ij}\bracketiii{\int\limits_{\mathcal{A}_{g-1,\vect{n}+2e_\alpha}} \omega^{k|l}_{g-1,\vect{n}+2e_\alpha} } \pt
\ee
The maps $\oosym{i}{j}$ and $\oxisym{ij}$ denote the contraction w.r.t. $\omega_\alpha^{-1}$, which is the inverse of the bpz inner product $\op{bpz}_\alpha^{-1}$ plus additional insertion originating from the sewing operations (\ref{eq:sewing}). In the following we determine these insertions.

In every sector $\alpha$ we have the twist angle $\vartheta$, which leads to an insertion
\be\label{eq:insertionvartheta}
\int_0^{2\pi}d\vartheta\,  B(v_\vartheta) \exp(i\vartheta L_0^{-}) \com
\ee
where $\exp(i\vartheta L_0^{-})$ generates the twisting and $B(v_\vartheta)$ originates from the measure (\ref{eq:forms}). The vector $v_\vartheta$ is determined by
\be\no
\pd_\vartheta \exp(i\vartheta L_0^{-})=iL_0^-\exp(i\vartheta L_0^{-}) \quad \Rightarrow \quad v_\vartheta= il_0^- \pt
\ee
In the case of R punctures in the holomorphic sector, we have the additional odd parameter $\tau$. Consequently, the corresponding Virasoro vector is odd and the measure contributes a picture changing operator. The insertion associated to $\tau$ reads
\be\label{eq:insertiontau}
\int d\tau \, \delta(B(v_\tau))\exp(\tau G_0) \pt
\ee
From
\be\no
\pd_\tau \exp(\tau G_0)=(G_0+\tau L_0)\exp(\tau G_0)
\ee
we conclude that
\be\no
v_\tau=g_0+\tau l_0 \pt
\ee
Combining (\ref{eq:insertionvartheta}) and (\ref{eq:insertiontau}) and carrying out the integrals using some of the identities of appendix \ref{app:forms}, we end up with \cite{Yeh}
\begin{align}\label{eq:sym}
\omega_{NS-NS}^{-1}&=2\pi i b_0^- P_{L_0^-} \circ \op{bpz}_{NS-NS}^{-1}\\\no
\omega_{R-NS}^{-1}&=2\pi i b_0^- P_{L_0^-} X_{g_0} \circ \op{bpz}_{R-NS}^{-1}\\\no
\omega_{NS-R}^{-1}&=2\pi i b_0^- P_{L_0^-} \ti{X}_{\ti{g}_0} \circ \op{bpz}_{R-NS}^{-1}\\\no
\omega_{R-R}^{-1}&=2\pi i b_0^- P_{L_0^-} X_{g_0} \ti{X}_{\ti{g}_0} \circ \op{bpz}_{R-R}^{-1} \pt
\end{align}
In equation (\ref{eq:sym}), we think of $\op{bpz}_\alpha^{-1}$ as a map from the dual space $\mathcal{H}^\ast_\alpha$ to $\mathcal{H}_\alpha$, and $P_{L_0^-}$ denotes the projection onto states satisfying the level matching condition. Moreover, the operator
\be\no
X_{g_0}= \inv{2}\bracket{G_0\delta(\beta_0)-\delta(\beta_0)G_0}
\ee
is the picture changing operator associated to $g_0$ (see appendix \ref{app:forms}). In the following we will discard the factor of $2\pi i$ in (\ref{eq:sym}), which has to be compensated by the normalization $N_{g,\vect{n}}$ introduced for the differential forms (\ref{eq:forms}). The restricted state space $\hat{\mathcal{H}}_\alpha$ is now determined by demanding that $\omega^{-1}_\alpha$ is indeed the inverse of a map $\omega_\alpha:\hat{\mathcal{H}}\to \hat{\mathcal{H}}^\ast$, the odd symplectic form relevant for the BV formalism. The constraint shared in all sectors is the level matching condition and $[Q,L_0^-]=b_0^-=0$. Consider now the holomorphic R sector. Since $\beta_0^2 X_{g_0}=0$, we conclude that states in the corresponding restricted state space have to satisfy
\be\no
\beta_0^2 =0 \pt
\ee
Furthermore, gauge invariance requires also
\be\no
\inv{2}[Q,\beta_0^2]=G_0\beta_0-b_0=0 \pt
\ee
In table \ref{tab:restrictions}, we summarize the constraints defining the restricted state spaces in the various sectors.

\begin{table}
\begin{center}
\begin{tabular}{c|c|c|c}
NS-NS & R-NS & NS-R & R-R \\\hline\hline
$\qquad L_0^-=0\qquad$ & $\qquad L_0^-=0 \qquad$ & $\qquad L_0^-=0 \qquad$ & $\qquad L_0^-=0 \qquad$ \\
 $b_0^-=0$ & $b_0^-=0$ & $b_0^-=0$ & $b_0^-=0$ \\
& $\beta_0^2=0$ & $\ti{\beta}_0^2=0$ & $\beta_0^2=\ti{\beta}_0^2=0$ \\
& $G_0\beta_0-b_0=0$ & $\ti{G}_0\ti{\beta}_0-\ti{b}_0=0$ & $\ti{G}_0\ti{\beta}_0-\ti{b}_0=G_0\beta_0-b_0=0$
\end{tabular}\caption{Constraints defining restricted state spaces.}\label{tab:restrictions}
\end{center}
\end{table}

An odd symplectic form is by definition an antisymmetric, closed, non-degenerate, bilinear map. But note that symmetry properties depend on the choice of grading. Consider for example a symmetric bilinear map $g:V^{\tp2}\to \C$ on a graded vector space $V=\ds_n V_n$. The suspension map $\uparrow$ and the desuspension map $\downarrow$ are defined by $(\uparrow V)_n=V_{n-1}$ and $(\downarrow V)_n=V_{n+1}$. The map $g\circ (\downarrow \tp \downarrow): \,\uparrow V^{\tp2}\to \C$ induced on $\uparrow V$ defines then an antisymmetric map. From a mathematical point of view, the natural choice of grading in string field theory is determined by declaring the degree of a classical field to be zero. Of course this does not coincide with the ghost number, picture and Grassmann parity grading defined in appendices \ref{app:scft} and \ref{app:forms}: From the requirement that on-shell amplitudes are of degree $0|0|0$, which are defined by integrating $\omega^{0|0}_{g,\vect{n}}$ over the full moduli space ${\mathfrak{M}}^{II}_{g,\vect{n}}\,$, we infer from (\ref{eq:degreeforms}) that the degrees of classical fields are given by table \ref{tab:degreefields}.

\begin{table}
\begin{center}
\begin{tabular}{c|c|c|c}
NS-NS & R-NS & NS-R & R-R \\\hline
$\quad 2|2|0 \quad$ & $\quad2|1|1\quad$ & $\quad2|1|1\quad$ & $\quad2|0|0\quad$
\end{tabular}\caption{Degrees of classical fields.}\label{tab:degreefields}
\end{center}
\end{table}

For every sector separately, we define a new grading
\be\label{eq:gradingnew}
g^\prime\|l^\prime\|\alpha^\prime\defineL g|p|\alpha -\text{degree of classical fields}\com
\ee
which sets the degree of classical fields to zero. We denote the corresponding desuspended space of $\hat{\mathcal{H}}_\alpha$ by $A_\alpha$. On $A_\alpha$ the odd symplectic form, which we also denote by $\omega_\alpha$, reveals its natural properties, i.e. it is antisymmetric and of degree $-1\|0\|0$\footnote{Note that a suspension/desuspension in picture does not change the symmetry properties in contrast to ghost number and Grassmann parity.}. $\omega_\alpha$ is the composite of the bpz inner product and an insertion which is inverse to the insertion of (\ref{eq:sym}). The insertion has to be BRST closed and has to have the appropriate bpz parity\footnote{Since the bpz inner product is symmetric on $\mathcal{H}_\alpha$, $\omega_\alpha$ being antisymmetric requires a bpz odd/even insertion and an even/odd number of desuspensions in ghost number and Grassmann parity.}. The symplectic forms, expressed via the bpz inner product and the additional insertion read \cite{Yeh}
\begin{align}\label{eq:symform}
\omega_{NS-NS} &= \op{bpz}_\alpha (\cdot, c_0^-\cdot) \\\no
\omega_{R-NS} &= \op{bpz}_\alpha \bracketi{\,\cdot\,,-2c_0^-c_0^+ \delta^\prime(\gamma_0)\,\cdot\,} \\\no
\omega_{NS-R} &= \op{bpz}_\alpha  \bracketi{\,\cdot\,,-2c_0^-c_0^+ \delta^\prime(\ti{\gamma}_0)\,\cdot\,} \\\no
\omega_{R-R} &= \op{bpz}_\alpha\bracketi{\,\cdot\,,-c_0^- G_0^{-1}\ti{G}_0^{-1} \delta(\gamma_0)\delta(\ti{\gamma}_0)\,\cdot\,} \pt
\end{align}

In the restricted state space, we can impose the Siegel gauge conditions \cite{Thorn} as depicted in table \ref{tab:siegelgauge}.

\begin{table}
\begin{center}
\begin{tabular}{c|c|c|c}
NS-NS & R-NS & NS-R & R-R \\\hline
$\quad b_0^+=0 \quad$ & $\quad \beta_0=0 \quad$ & $\quad \ti{\beta}_0=0 \quad$ & $\quad b_0^+=0\sep (G_0\neq0\neq \ti{G}_0)  \quad$
\end{tabular}\caption{Siegel gauge in the various sectors.}\label{tab:siegelgauge}
\end{center}
\end{table}

Surprisingly, the symplectic form in the R-R sector is non-local and degenerates on-shell, but in section \ref{sec:vertices} we will see that in combination with the BRST charge, this will reproduce the right expression for the propagator.

Finally, similarly to the geometric BV structure described in section \ref{sec:geobv}, we can define a BV structure on
\be\label{eq:hominv}
\op{Hom}_{\op{inv}}(A^{\tp\vect{n}},\C^{1|1})\defineL \bracketii{\op{Hom}(A^{\tp\vect{n}},\C^{1|1})\tp\oMod(\oCom^{\mathcal{N}=1})(g,\vect{n})}^{\Sigma_{\vect{n}}} \pt
\ee
The antibracket is defined by
\begin{align}\label{eq:algop}
(h_{g_1,\vect{n}_1+e_\alpha},h_{g_2,\vect{n}_2+e_\alpha})^{\op{alg}}_\alpha &\defineL  \sum_{\sigma\in\op{sh}\bracket{\vect{n}_1,\vect{n}_2}} \sigma.\bracketi{h_{g_1,\vect{n}_1+e_\alpha} \,\oosym{i}{j}\, h_{g_2,\vect{n}_2+e_\alpha}} \com \\\no
(\cdot,\cdot)^{\op{alg}} &\defineL \sum_\alpha (\cdot,\cdot)^{\op{alg}}_\alpha \pt
\end{align}
and the BV operator reads
\begin{align}\label{eq:algbv}
\Delta^{\op{alg}}_\alpha h_{g,\vect{n}+2e_\alpha}  &\defineL  \oxisym{ij} \bracket{h_{g,\vect{n}+2e_\alpha}}    \com\\\no
\Delta^{\op{alg}} &\defineL \sum_\alpha\Delta^{\op{alg}}_\alpha \pt
\end{align}
for $h_{g,\vect{n}+2e_\alpha}\in \op{Hom}_{\op{inv}}(A^{\tp\vect{n}+2e_\alpha},\C^{1|1})$, $h_{g,\vect{n}_i+e_\alpha}\in \op{Hom}_{\op{inv}}(A^{\tp\vect{n}_i+e_\alpha},\C^{1|1})$. The permutation $\sigma$ in equation (\ref{eq:algop}) acts by permuting the inputs of the linear map.

From the factorization properties (\ref{eq:factorizationsym1}), (\ref{eq:factorizationsym2}) and the chain map property (\ref{eq:chainmapproperty}), we infer that the STCFT defines a morphism of BV algebras, i.e.
\be\no
\op{STCFT}: \bracketii{C^{\bullet|\bullet}_{\op{inv}}(\hat{\mathfrak{P}}^{II}_{g,\vect{n}}), \,\pd, \,\Delta^{\op{geo}},\,(\cdot,\cdot)^{\op{geo}}} \to
\bracketii{\op{Hom}_{\op{inv}}(A^{\tp\vect{n}},\C^{1|1}), \,Q, \,\Delta^{\op{alg}},\,(\cdot,\cdot)^{\op{alg}}} \pt
\ee

\section{Vertices and BV Master Equation}\label{sec:vertices}
In this part we construct the vertices for type II super string field theory. First, we discuss the kinetic term, and in particular its form in Siegel gauge. In a second step we treat the interactions and show that a consistent decomposition of the moduli space implies that the vertices satisfy a BV master equation. Finally, we outline an explicit construction of the vertices in close analogy to the bosonic case \cite{Zwiebach closed}, by formulating a minimal area problem for type II world sheets.

\subsection{Kinetic term}
The kinetic term for a string field $\phi_\alpha\in A_\alpha$ of degree $0\|0\|0$ is defined by
\be
\omega_\alpha(Q\phi_\alpha,\phi_\alpha) \pt
\ee
In Siegel gauge (see table \ref{tab:siegelgauge}) the kinetic term reduces to
\begin{align}\label{eq:propsiegel}
\omega_{NS-NS}\bracketi{L_0^+ c_0^+ \phi,\phi}&=\op{bpz}_{NS-NS}\bracketi{c_0^-c_0^+ L_0^+ \phi,\phi}\sep \phi\in A_{NS-NS}\\\no
\omega_{R-NS}\bracketi{G_0\gamma_0 \phi,\phi}&=\op{bpz}_{R-NS}\bracketi{-2c_0^-c_0^+ \delta(\gamma_0) G_0  \phi,\phi}\sep \phi\in A_{R-NS}\\\no
\omega_{NS-R}\bracketi{\ti{G}_0\ti{\gamma}_0 \phi,\phi}&=\op{bpz}_{NS-R}\bracketi{-2c_0^-c_0^+ \delta(\ti{\gamma}_0) \ti{G}_0  \phi,\phi}\sep \phi\in A_{NS-R}\\\no
\omega_{R-R}\bracketi{L_0^+c_0^+ \phi,\phi}&=\op{bpz}_{R-R}\bracketi{-c_0^-c_0^+\delta(\gamma_0)\delta(\ti{\gamma}_0) G_0^{-1}\ti{G}_0^{-1}L_0^+  \phi,\phi}\sep \phi\in A_{R-R}\pt
\end{align}

The insertions in the bpz inner product of equation (\ref{eq:propsiegel}) lead precisely to the propagators known from perturbative string theory \cite{Witten perturbation}. We conclude that the non-local form of the kinetic term in the R-R sector is probably related to the problem of finding an action principle for a self dual field strength.

\subsection{Interactions}
The covariant kinetic term defined in the previous subsection requires intrinsically a background. In contrast, the interactions represent a subspace of the moduli space. We call the corresponding vertices the \emph{geometric vertices}. In order to be consistent with perturbative string theory, the geometric vertices have to reproduce a single cover of the full moduli space via Feynman rules. For a given background, which determines a TSCFT, the image of the geometric vertices under the TSCFT defines the corresponding \emph{algebraic vertices}. Thus the geometric vertices are background independent, whereas the algebraic vertices depend on the choice of background.

To formulate the consistency condition for the geometric vertices, we first have to define the notion of propagation on moduli space: The \emph{geometric propagator} is defined by sewing of punctures w.r.t.
\begin{align}\label{eq:geoprop}
(P_{NS-NS})_{x,\vartheta}&=\bracketi{ I_{NS}\circ \varphi^{l_0}_{-x+i\vartheta} \,,\, \ti{I}_{NS}\circ \varphi^{\ti{l}_0}_{-x-i\vartheta}} \\\no
(P_{R-NS})_{x,\vartheta,\tau}&=\bracketi{I_{R}\circ \varphi^{g_0}_{-x+i\vartheta,\tau} \,,\, \ti{I}_{NS}\circ \varphi^{\ti{l}_0}_{-x-i\vartheta}}  \\\no
(P_{NS-R})_{x,\vartheta,\ti{\tau}}&=\bracketi{I_{NS}\circ \varphi^{{l}_0}_{-x+i\vartheta}  \,,\, \ti{I}_{R}\circ \varphi^{\ti{g}_0}_{-x-i\vartheta,\ti{\tau}}}  \\\no
(P_{R-R})_{x,\vartheta,\tau,\ti{\tau}}&=\bracketi{I_{R}\circ \varphi^{{g}_0}_{-x+i\vartheta,\tau}  \,,\, \ti{I}_{R}\circ \varphi^{\ti{g}_0}_{-x-i\vartheta,\ti{\tau}}} \com
\end{align}
for $x\in [0,\infty)$, $\vartheta\in[0,2\pi]$ and $\tau,\ti{\tau}\in \C^{0|1}$. The quantity $x$ can be interpreted as the length of the cylinder sewn in between two puncture, and the sewing maps defined in equation (\ref{eq:sewing}) correspond to setting $x=0$. The induced maps on the invariant chain complex $C^{\bullet|\bullet}_{\op{inv}}(\hat{\mathfrak{P}}^{II}_{g,\vect{n}})$ carry degree $0|0$.

The geometric vertices $\mathcal{V}_{g,\vect{n}}$ represent a subspace of codimensionality $0|0$ of the moduli space decorated with coordinate curves, invariant under permutation of punctures of the same type. In other words,  $\mathcal{V}_{g,\vect{n}}\in C^{0|0}_{\op{inv}}(\hat{\mathfrak{P}}^{II}_{g,\vect{n}})$. From the collection of geometric vertices, we can construct graphs with the aid of the propagator\footnote{In section \ref{sec:operad}, while introducing operads, we will state more precisely what we mean by graphs.}. We denote the collection of genus $g$ graphs with $\vect{n}$ punctures, constructed from $\mathcal{V}_{g,\vect{n}}$ and involving exactly $i$ propagators, by $R^i_{g,\vect{n}}$. The requirement of a single cover reads \cite{Zwiebach closed}
\be\label{eq:singlecover}
\overline{\mathfrak{M}}^{II}_{g,\vect{n}}= \pi\bracketi{\mathcal{V}_{g,\vect{n}} \sqcup R^1_{g,\vect{n}} \sqcup\dots\sqcup R^{3g-3+n}_{g,\vect{n}}} \com
\ee
where $3g-3+n$ is the maximal possible number of propagators, $\overline{\mathfrak{M}}^{II}_{g,\vect{n}}$ denotes the Deligne-Mumford compactification of ${\mathfrak{M}}^{II}_{g,\vect{n}}$ \cite{Witten srs} and $\pi$ denotes the projection map on $\hat{\mathfrak{P}}^{II}_{g,\vect{n}}$ as a fibre bundle over $\mathfrak{M}^{II}_{g,\vect{n}}$. The degenerations arise from infinitely long cylinders, i.e. correspond to $x\to \infty$.

The compactified moduli space on the left hand side of equation (\ref{eq:singlecover}) has no boundary. On the other hand the right hand side of equation (\ref{eq:singlecover}) involves two types of boundaries: One which describes the boundary of the geometric vertices itself and another which corresponds to a propagator collapse, i.e. $x\to0$. Thus equation (\ref{eq:singlecover}) implies that these two types of boundaries cancel each other. The required canellation of boundary terms is equivalent to the BV master equation \cite{Zwiebach closed}
\be\label{eq:geobvmaster}
\pd \mathcal{V}_{g,\vect{n}}+ \sum_\alpha\Delta^{\op{geo}}_\alpha \mathcal{V}_{g-1,\vect{n}+2e_\alpha}
+ \inv{2}\sum_\alpha\sum_{\vect{n}_1+\vect{n}_2=\vect{n} \atop g_1+g_2=g}( \mathcal{V}_{g_1,\vect{n}_1+e_\alpha}, \mathcal{V}_{g_2,\vect{n}_2+e_\alpha})^{\op{geo}}_\alpha=0 \pt
\ee
To summarize, every consistent decomposition of the moduli space into vertices and graphs implies that the BV master equation (\ref{eq:geobvmaster}) is satisfied.

In the rest of this section, we will introduce minimal area metrics on type II world sheets and outline their relevance for the construction of the geometric vertices. Following \cite{Witten srs}, a metric on a type II world sheet $\B{\Sigma}\subset\Sigma\times\ti{\Sigma}$ is determined by a collection of even local sections
\begin{align}\no
E\in \,&\Gamma(U,\D^{-2}) \com\\\no
\ti{E}\in \,&\Gamma(U,\ti{\D}^{-2})
\end{align}
where $\D$ denotes the distinguished subbundle of $T\B{\Sigma}$ (see appendix \ref{app:srs}). Overlapping sections are related by the gauge transformation
\begin{align}\no
E^\prime&=e^{iu}E \com\\\no
\ti{E}^\prime&=e^{-iu}\ti{E} \com
\end{align}
satisfying the reality condition $\overline{E}=\ti{E}$ and $u \in \R$ for $\vt=\ti{\vt}=0$. The subbundle $\D\subset T{\Sigma}$ is locally spanned by $D_\vt=\pd_\vt+\vt \pd_z$, whereas $\D^{-2}\subset T^\ast\Sigma$ is locally spanned by $\Omega_z=dz+\vt d\vt$. Hence, $(D_\vt,\pd_z)$ describes a basis of $T\Sigma$ respecting the superconformal structure, with dual basis $(d\vt,\Omega_z)$. For a given coordinate system $(z,\vt)$, a local section $E$/$\ti{E}$ determines $\varphi$/$\ti{\varphi}$ via
\begin{align}\no
E&=e^\varphi \Omega_z \com\\\no
\ti{E}&= e^{\ti{\varphi}}\Omega_{\ti{z}} \pt
\end{align}
Furthermore, there is an odd one-form $F$/$\ti{F}$ determined (up to a sign) by
\begin{align}\label{eq:F}
\pi(dE)=F\w F \com \\\no
\ti{\pi}(d\ti{E})=\ti{F}\w \ti{F} \com
\end{align}
where $\pi$/$\ti{\pi}$ denotes the projection maps onto $T^\ast\Sigma\tp T^\ast\Sigma$/$T^\ast\ti{\Sigma}\tp T^\ast\ti{\Sigma}$.
From equation (\ref{eq:F}), we infer
\begin{align}\no
F&=e^{\varphi/2}\bracketi{d\vt+\tinv{2}D_\vt\varphi\Omega_z} \com\\\no
\ti{F}&=e^{\ti{\varphi}/2}\bracketi{d\ti{\vt}+\tinv{2}D_{\ti{\vt}}\ti{\varphi}\Omega_{\ti{z}}} \pt
\end{align}
The full metric $G$, globally defined on $\B{\Sigma}$, then reads
\be\no
G= E\tp \ti{E}+\ti{E}\tp E +F\tp \ti{F}-\ti{F}\tp F \pt
\ee
The area of $\B{\Sigma}$ measured w.r.t. the metric $G$ is defined by
\be\label{eq:area}
A(\B{\Sigma})=\int dz d\ti{z} d\vt d\ti{\vt} \bracketi{\op{sdet}({}_iG_j)}^{1/2} \pt
\ee
Here we use the left and right index notion introduced in \cite{Witt super}. It can be shown \cite{Witten srs}, that the superdeterminant bundle $\op{sdet}(\Sigma)$ is isomorphic to $\D^{-1}$. Thus a volume form for a type II world sheet naturally defines a section of $\D^{-1}\tp\ti{\D}^{-1}$. By a straightforward calculation, one can verify that
\be\no
\bracketi{\op{sdet}({}_iG_j)}^{1/2}=e^{(\varphi+\ti{\varphi})/2} \com
\ee
i.e. that it transforms as a section of $\D^{-1}\tp\ti{\D}^{-1}$.
Finally, consider a supercircle $\gamma:S_\alpha^{1|2}\to \B{\Sigma}$ embedded in $\B{\Sigma}$. The length of $\gamma$ measured with the induced metric reads
\be\no
L(\gamma)=\int dt d\tau d\ti{\tau}\bracketi{\op{sdet}({}_i(\gamma^\ast G)_j)}^{1/2} \pt
\ee

Now we have all the necessary ingredients to formulate the appropriate minimal area problem: For a given type II world sheet $\B{\Sigma}$, we ask for the metric of minimal area under the condition that there is no non-trivial supercircle which is shorter than $2\pi$. We conjecture that this minimal area problem has a unique solution.

In analogy to the bosonic case \cite{Zwiebach closed}, we claim that a minimal area metric on $\B{\Sigma}$ gives rise to \emph{bands of saturating geodesics}: A saturating geodesic is a supercircle whose length is exactly $2\pi$. Furthermore, saturating geodesics of the same homotopy class are non-intersecting. The collection of all saturating geodesics of a certain homotopy class foliate a part of $\B{\Sigma}$, which is called a band of saturating geodesics. Note that in general bands of saturating geodesics might intersect.

A band of saturating geodesics has the topology of a supercylinder. The height of a band of saturating geodesics is defined to be the shortest superpath between the two boundary components. We distinguish external bands from internal bands, by whether the saturating geodesics are homotopic to a puncture or not.

An external band describes a semi-infinite supercylinder, that is there is a bounding saturating geodesic from where the band extends infinitely towards the puncture. We can now define a section
\be\label{eq:section}
\sigma^{l}: \mathfrak{M}^{II}_{g,\vect{n}} \to \hat{\mathfrak{P}}^{II}_{g,\vect{n}}\com
\ee
by defining coordinate curves to be the saturating geodesic a distance $l$ separated from the bounding saturating geodesic.
The smallest possible choice for $l$ is $\pi$, since for $l\le \pi$ the sewing of two punctures would lead to supercircles shorter than $2\pi$.

Finally, we describe a 1-parameter family of vertices satisfying condition (\ref{eq:singlecover}) \cite{Zwiebach closed}:  For given $l\ge\pi$, we define $\mathcal{U}^{l}_{g,\vect{n}}$ to be the collection of surfaces $\B{\Sigma}\in\mathfrak{M}^{II}_{g,\vect{n}}$, which have no internal bands of saturating geodesics of height larger than $l$. The vertices together with coordinate curves are then defined by
\be\label{eq:geovertices}
\mathcal{V}^{l}_{g,\vect{n}}\defineL \sigma^{l}\bracket{\mathcal{U}^{l}_{g,\vect{n}}} \in C^{0|0}_{\op{inv}}(\hat{\mathfrak{P}}^{II}_{g,\vect{n}})  \pt
\ee
According to (\ref{eq:geobvmaster}), the BV master equation
\be\label{eq:algbvmaster}
\pd \mathcal{V}^l_{g,\vect{n}}+ \sum_\alpha\Delta^{\op{geo}}_\alpha \mathcal{V}^l_{g-1,\vect{n}+2e_\alpha}
+ \inv{2}\sum_\alpha\sum_{\vect{n}_1+\vect{n}_2=\vect{n} \atop g_1+g_2=g}( \mathcal{V}^l_{g_1,\vect{n}_1+e_\alpha}, \mathcal{V}^l_{g_2,\vect{n}_2+e_\alpha})^{\op{geo}}_\alpha =0\com
\ee
is satisfied.

From a field theory point of view, the parameter $l-\pi$ can be interpreted as a cut-off. There are two interesting limits: The vertices corresponding to $l\to \pi$ describe the smallest possible subset of the moduli space consistent with (\ref{eq:singlecover}). This is the natural choice of geometric vertices. On the other hand, in the limit $l\to \infty$ we have $\mathcal{U}^{l}_{g,\vect{n}}=\mathfrak{M}^{II}_{g,\vect{n}}$, and the corresponding master equation describes the Deligne-Mumford compactification. Indeed, in this singular limit the assignment of coordinate curves is obsolete \cite{Witten srs}, and thus the master equation describing the compactification can actually be formulated without a global section $\sigma:\mathfrak{M}^{II}_{g,\vect{n}} \to \hat{\mathfrak{P}}^{II}_{g,\vect{n}}$.

For a given TSCFT (background), the corresponding algebraic vertices $f_{g,\vect{n}}\in \op{Hom}_{\op{inv}}(A^{\tp\vect{n}},\C^{1|1})$ are now defined by
\be\label{eq:algvertices}
f_{g,\vect{n}}= \int_{\mathcal{V}_{g,\vect{n}}} \omega^{0|0}_{g,\vect{n}} \pt
\ee
Since the TSCFT defines a morphism of BV algebras (see section \ref{sec:algbv}), the algebraic vertices satisfy the BV master equation
\be\no
f_{g,\vect{n}}\circ \sum_{i=1}^n Q^{(i)}+ \sum_\alpha\Delta^{\op{alg}}_\alpha f_{g-1,\vect{n}+2e_\alpha}
+\inv{2}\sum_\alpha\sum_{\vect{n}_1+\vect{n}_2=\vect{n}\atop g_1+g_2=g}(f_{g_1,\vect{n}_1+e_\alpha},f_{g_1,\vect{n}_1+e_\alpha})^{\op{alg}}_\alpha =0 \pt
\ee
The relevant grading for the BV formalism is the ghost number and the Grassmann parity, but not the picture. That is, the picture number of fields and antifields coincides with the picture number of classical fields (see table \ref{tab:degreefields}). Fields have ghost number less then or equal to the ghost number of classical fields, and alternate in Grassmann parity. Similarly, antifields have ghost number greater then classical fields and alternate in Grassmann parity as well. In other words, we restrict the two outputs of the inverse of the symplectic structure appearing in the antibracket and the BV operator to the picture number of classical fields.

Finally, the full quantum action satisfying the BV master equation reads
\be\label{eq:action}
S(\B{c})=\inv{2}\sum_\alpha \omega_\alpha(Qc_\alpha,c_\alpha) + \sum_{g,\vect{n}}\frac{\hbar^g}{\prod_\alpha n_\alpha!}\,f_{g,\vect{n}}(\B{c}^{\vect{n}}) \com
\ee
where $\B{c}=(c_\alpha)$ denotes the collection of fields and antifields in the various sectors.

\section{Algebraic Structure and Operadic Description}\label{sec:operad}
In this section, we employ operads in order to restate the result of the previous section in a uniform and concise way. It will turn out that the construction of string field theory can be formulated by two morphisms between appropriate modular operads, one which describes the decomposition of the moduli space and a second which represents the background.

We start with a brief introduction of modular operads and the Feynman transform. We then quote a result of \cite{Barannikov operadBV}, which establishes a relation between algebras over the Feynman transform of a modular operad and solutions to a corresponding BV master equation. This introductory part does not claim full mathematical rigor, but is rather intended to develop some intuition. We refer the interested reader to \cite{Getzler modop, Markl operads, Barannikov operadBV} for a thorough exposition.

A stable $\Sigma$-module $\oP$ is a collection of differential graded vector spaces $\oP(g,n)$ endowed with a $\Sigma_n$ action, for all $g\ge0$ and $n\ge0$ satisfying the stability condition $2g+n-3\ge0$.

A graph $G$ is a collection $(H(G),V(G),\pi,\sigma)$, where the half-edges $H(G)$ and the vertices $V(G)$ are finite sets, $\pi:H(G)\to V(G)$ and $\sigma:H(G)\to H(G)$ is an involution, i.e. $\sigma^2=\op{id}$.

The preimage $\pi^{-1}(v)\defineR L(v)$ determines the half-edges attached to the vertex $v\in V(G)$. The cardinality of $L(v)$ is denoted by $n(v)$. The involution $\sigma$ decomposes into 1-cycles and 2-cycles, where the 1-cycles define the legs (external lines) $L(G)$ and the 2-cycles define the edges (internal lines) $E(G)$ of the graph $G$.

A stable graph is a connected graph $G$ together with a map $g:V(G)\to \N_0$, which assign a genus to each vertex. For every vertex $v\in V(G)$ the stability condition $2g(v)+n(v)-3\ge0$ has to hold. The genus of the graph $G$ is defined by $g(G)=\sum_{v\in V(G)}g(v)+ b_1(G)$, where $b_1(G)$ denotes the first Betti number. Furthermore we require a bijection between $L(G)$ and $\{1,\dots,n(G)\}$, where $n(G)$ denotes the cardinality of $L(G)$.

A morphism of graphs is a contraction of edges. Let $G$ be a stable graph and $I\subset E(G)$ a subset of its edges. We denote the graph that arises from contracting the edges $I$ of the graph $G$ by $G/I$,
and the corresponding morphism by $f_{G,I}:G\to G/I$. Every morphism can be decomposed into a collection of single edge contraction. There are two types of single edge contractions, corresponding to the separating and non-separating case, i.e. to the contraction of an edge connecting two vertices and the contraction of an edge forming a loop on one vertex respectively. In the following, we use a graphical representation for the single edge graphs
\be\no
\Tree
\ee
and
\be\no
\hspace{0.6cm}\Loop
\ee
in the separating and non-separating case respectively. Stable graphs and morphism as described above define the category $\Gamma(g,n)$.

Let $\oP$ be a stable $\Sigma$-module and $G$ a stable graph. We define
\be\no
\oP(G)=\bigotimes_{v\in V(G)}\oP(g(v),n(v)) \pt
\ee
A \emph{modular operad} $\oP$ is a stable $\Sigma$-module, which in addition defines a functor on the category of graphs. That is, for every morphism $f:G_1\to G_2$ there is a morphism $\oP(f):\oP(G_1)\to \oP(G_2)$, and the associativity condition
\be\no
\oP(f\circ g)=\oP(f)\circ\oP(g)
\ee
has to hold. A \emph{cyclic operad} is the tree level version of a modular operad, i.e. corresponds to $g=0$.

Due to the functor property and the fact that every morphism of graphs can be decomposed into single edge contractions, a modular operad $\oP$ is indeed determined by the underlying $\Sigma$-module together with the maps
\be\no
\oP\bracketi{f_{\Etree}}\defineR \oo{i}{j}
\ee
and
\be\no
\oP\bracketi{f_{\Eloop}}\defineR \oxi{ij} \com
\ee
where $i$ and $j$ represent the half edges constituting the edge $e$.

Finally, there is the notion of twisted modular operads. The only twist we will need is the so called $\mathfrak{K}$-twist, which assigns degree one to the edges of a graph: For a stable graph $G$, $\mathfrak{K}(G)$ is defined to be the top exterior power of the vector space generated by the elements of $E(G)=\{e_1,\dots,e_n\}$, suspended to degree $n$, i.e.
\be\no
\mathfrak{K}(G)=\op{det}(E(G))\defineL \, \uparrow^n \Lambda^n \bracketi{\op{span}(E(G))} \pt
\ee

The standard example of a modular operad is the endomorphism operad. Let $(A,d)$ be a differential graded vector space endowed with a symmetric, bilinear and non-degenerate form $B:A^{\tp 2}\to\Bbbk$ of degree zero, where $\Bbbk$ denotes some field or ring. The inverse $B^{-1}$ of $B$ is also symmetric and of degree zero. We define the $\Sigma_n$-modules
\be\no
\mathcal{E}[A,d,B](g,n)=\op{Hom}(A^{\tp n},\Bbbk) \com
\ee
where the action of $\Sigma_n$ is defined by permutation of the inputs of the multilinear maps. Contractions w.r.t. $B^{-1}$ make $\mathcal{E}[A,d,B]$ a modular operad.
Similarly, consider a differential graded vector space $(A,d)$ endowed with an odd symplectic structure of degree $-1$. The inverse $\omega^{-1}$ is then symmetric and of degree $1$. Due to the degree of $\omega^{-1}$,
\be\no
\mathcal{E}[A,d,\omega](g,n)=\op{Hom}(A^{\tp n},\Bbbk)
\ee
defines a $\mathfrak{K}$-twisted modular operad.

An algebra over a modular operad $\oP$, called a $\oP$-algebra, is a morphism $\alpha$ form $\oP$ to some endomorphism operad.

The last ingredient we need is the Feynman transform of a modular operad. Let $\M$ be the functor from the category of stable $\Sigma$-modules to the category of modular operads, left adjoint to the forgetful functor. Consider a modular operad $\oP$ and let $\oP(g,n)^\ast$ be the dual space of $\oP(g,n)$. For our purposes, it suffices to consider the case where the differential on $\oP$ vanishes, i.e. $d_{\oP}=0$. The Feynman transform $\oF\oP$ of $\oP$ is defined to be the $\mathfrak{K}$-twisted modular operad freely generated from the dual spaces $\oP(g,n)^\ast$, i.e.
\be\no
\oF\oP= \M_\mathfrak{K}\oP^\ast\defineL \bigoplus_{G\in [\Gamma(g,n)]}\bracketi{\mathfrak{K}(G)\tp \oP(G)^\ast}_{\op{Aut}(G)} \com
\ee
where $[\Gamma(g,n)]$ denotes the set of isomorphism classes of stable graphs. The main feature of the Feynman transform is that it endows $\oF\oP$ with an additional differential: The \emph{Feynman differential} $d_{\oF\oP}$ is defined by
\be\no
d_{\oF\oP} {\big |}_{\bracket{\mathfrak{K}(G) \tp \oP(G)^\ast}_{\op{Aut}(G)}}=\sum_{G^\prime/\{e\}\simeq G} \uparrow e\tp\oP(f_{G^\prime,\{e\}})^\ast \com
\ee
i.e. for a given graph $G$ it generates all graphs $G^\prime$ which are isomorphic to $G$ upon contracting a single edge $e$.

Consider now a morphism $\alpha$ from the Feynman transform $\oF\oP$ of a modular operad $\oP$ to some $\mathfrak{K}$-twisted modular operad $\oQ$. The morphism is $\Sigma$ equivariant and defines a chain map, i.e.
\be
d_{\oQ}\circ\alpha=\alpha\circ d_{\oF\oP} \pt
\ee
Furthermore, $\alpha$ is determined by
\be\label{eq:morphismFP}
\alpha(g,n):\oP(g,n)^\ast\to \oQ(g,n) \com
\ee
and $\Sigma_n$ equivariance implies that
\be\no
\alpha(g,n)\in \bracketi{ \oQ(g,n)\tp\oP(g,n)}^{\Sigma_n} \pt
\ee

Evaluating equation (\ref{eq:morphismFP}) on a graph consisting of a single vertex leads to \cite{Barannikov operadBV}
\begin{align}\label{eq:morphismFPex}
d_{\oQ}\circ \alpha(g,I)=\oQ\bracketi{f_{\Eloop}}\tp\oP\bracketi{f_{\Eloop}}\bracketi{\uparrow &e\tp \alpha(g-1,I\sqcup \{i,j\})} \\\no
+\inv{2}\sum_{I_1\sqcup I_2=I \atop g_1+g_2=g}\oQ\bracketi{f_{\Etree}}\tp\oP\bracketi{f_{\Etree}}\bracketi{\uparrow &e\tp \alpha(g_1,I_1\sqcup\{i\})\tp \alpha(g_2,I_2\sqcup\{j\})}\com
\end{align}
where $I=\{1,\dots,n\}$. Equation (\ref{eq:morphismFPex}) can be interpreted as a BV master equation on $\bracketi{ \oQ(g,n)\tp\oP(g,n)}^{\Sigma_n}$, by identifying the contractions w.r.t. $\oQ$ and $\oP$ together with the determinant of the edge as the antibracket $(\cdot,\cdot)$ in the separating and the BV operator $\Delta$ in the non-separating case. $d_{\oF\oP}^2=0$ is then equivalent to the axioms of a BV algebra (without multiplication) listed in equation (\ref{eq:axiomsbv}) \cite{Barannikov operadBV}. Substituting $d_{\oQ}\to -d_{\oQ}$, equation (\ref{eq:morphismFPex}) reads
\be\label{eq:mastereqFP}
d_{\oQ}\circ \alpha(g,n)+\Delta \alpha(g-1,n+2)+\inv{2}\sum_{n_1+n_2=n\atop g_1+g_2=g}(\alpha(g_1,n_1+1),\alpha(g_2,n_2+1))=0 \pt
\ee

\begin{Thm}[\cite{Barannikov operadBV}]\label{thm:1}
Morphisms from the Feynman transform $\oF\oP$ of a modular operad $\oP$ to a $\mathfrak{K}$-twisted modular operad $\oQ$ are in one-to-one correspondence with solutions to the BV master equation (\ref{eq:mastereqFP}).
\end{Thm}

In the previous sections we saw that the geometric approach to string field theory inevitably leads to a certain BV master equation that has to be satisfied.
Thus, the link between the Feynman transform and solutions to an associated BV master equation immediately reveals the relevance of modular operads in the context of string field theory.

In type II superstring field theory, we have four different sectors $\alpha\in \{NS-NS,R-NS,NS-R,R-R\}$. Thus we need a slight generalization of a modular operad which allows for several sectors, i.e. a ``colored" version of a modular operad. For our purposes, a ``colored" modular operad $\oP$ is a collection of differential graded vector spaces $\oP(g,\vect{n})$, $\vect{n}=(n_\alpha)_{\alpha\in C}$, $n_\alpha\in\N_0$, satisfying the stability condition $2g+\sum_{\alpha}n_\alpha -3 \ge0$, where $C$ denotes the set of colors. Half edges of a graph are labeled by a color, and only half edges of the same color can form an edge. Furthermore we are only allowed to permute half edges of the same color, i.e. $\oP(g,\vect{n})$ is a $\Sigma_{\vect{n}}$-module, where $\Sigma_{\vect{n}}=\times_{\alpha}\Sigma_{n_\alpha}$.

In the following we introduce the relevant operads for the formulation of type II superstring field theory in terms of morphisms of operads. The cyclic operad encoding the symmetries of the classical (genus zero) vertices is denoted by $\oCom^{\mathcal{N}=1}$. It is a colored operad with $C=\{NS-NS,R-NS,NS-R,R-R\}$, and $\oCom^{\mathcal{N}=1}(\vect{n})$ are one dimensional vector spaces of degree zero without differential. The permutation group $\Sigma_\vect{n}$ acts trivially on $\oCom^{\mathcal{N}=1}(\vect{n})$. Furthermore, on top of the stability condition we impose the following constraints:
\begin{align}\label{eq:constraint}
\nnn+\nnr&\in \N_0 \\\no
\nnn+\nrn &\in \N_0\\\no
\nrr+\nrn &\in 2\N_0\\\no
\nrr+\nnr &\in 2\N_0 \pt
\end{align}

Let $x_{\vect{n}}$ denote the element that generates the vector space $\oCom^{\mathcal{N}=1}(\vect{n})$. The single edge contraction is defined by
\be\no
\oCom^{\mathcal{N}=1}\bracketi{f^\alpha_{\Etree}}\bracketi{x_{\vect{n}_1+e_\alpha}\tp x_{\vect{n}_2+e_\alpha}} = x_{\vect{n}_1+\vect{n}_2} \pt
\ee
It turns out that $\oCom^{\mathcal{N}=1}$ is generated by the vector spaces with $\sum_\alpha n_\alpha=3$. Such an operad is called a \emph{quadratic} operad \cite{Markl operads}. For $\sum_\alpha n_\alpha=3$ there are five cases compatible with (\ref{eq:constraint})
\bi
\item[(i)] $\nnn=3$
\item[(ii)] $\nnn=1$, $\nrn=2$
\item[(iii)] $\nnn=1$, $\nnr=2$
\item[(iv)] $\nnn=1$, $\nrr=2$
\item[(iv)] $\nrn=1=\nnr$, $\nrr=1$,
\ei
which correspond to the five possible types of genus zero surfaces with three punctures.

Let $\oMod$ be the functor from the category of cyclic operads to the category of modular operads, left adjoint to the forgetful functor. Consider now the modular operad $\oMod(\oCom^{\mathcal{N}=1})$ associated to the cyclic operad $\oCom^{\mathcal{N}=1}$, which encodes the symmetry properties of the vertices to all order in $\hbar$. Again $\oMod(\oCom^{\mathcal{N}=1})(g,\vect{n})$ are one dimensional vector spaces endowed with the trivial action of $\Sigma_{\vect{n}}$, and the single edge contractions read
\begin{align}\no
\oMod(\oCom^{\mathcal{N}=1})\bracketi{f^\alpha_{\Etree}}\bracketi{x_{g_1,\vect{n}_1+e_\alpha}\tp x_{g_2,\vect{n}_2+e_\alpha}} &= x_{g_1+g_2,\vect{n}_1+\vect{n}_2} \com \\\no
\oMod(\oCom^{\mathcal{N}=1})\bracketi{f^\alpha_{\Eloop}}\bracketi{x_{g-1,\vect{n}+2e_\alpha}} &= x_{g,\vect{n}} \com
\end{align}
where $x_{g,\vect{n}}$ is the element that generates $\oMod(\oCom^{\mathcal{N}=1})(g,\vect{n})$.

Next, we define the $\mathfrak{K}$-twisted modular operad $C^{\bullet|\bullet}(\hat{\mathfrak{P}}^{II})$. Its underlying $\Sigma_{\vect{n}}$-modules are $C^{\bullet|\bullet}(\hat{\mathfrak{P}}^{II}_{g,\vect{n}})$ with grading as defined in section \ref{sec:geobv}, and the single edge contractions are defined by
\begin{align}\no
C^{\bullet|\bullet}(\hat{\mathfrak{P}}^{II})\bracketi{f^\alpha_{\Etree}}\bracketi{\mathcal{A}_{g_1,\vect{n}_1+e_\alpha}\sqcup\mathcal{A}_{g_2,\vect{n}_2+e_\alpha}}
&=\mathcal{A}_{g_1,\vect{n}_1+e_\alpha}\,\oophi{i}{j}\,\mathcal{A}_{g_2,\vect{n}_2+e_\alpha}  \\\no
C^{\bullet|\bullet}(\hat{\mathfrak{P}}^{II})\bracketi{f^\alpha_{\Eloop}}\bracketi{\mathcal{A}_{g-1,\vect{n}+2e_\alpha}}
&=\oxiphi{ij}\mathcal{A}_{g-1,\vect{n}+2e_\alpha}  \com
\end{align}
where $\oophi{i}{j}$ and $\oxiphi{ij}$ are the sewing maps of equation (\ref{eq:sewingtree}) and equation (\ref{eq:sewingloop}) respectively.

Finally, consider a TSCFT which determines the endomorphism operad $\mathcal{E}[A_\alpha,Q_\alpha,\omega_\alpha]$, where $A_\alpha$ denotes the restricted state space with the grading of equation (\ref{eq:gradingnew}), $Q_\alpha$ is the BRST charge and $\omega_\alpha$ is the odd symplectic structure as defined in (\ref{eq:symform}).

As discussed in section \ref{sec:vertices}, a consistent decomposition of the moduli space into vertices and graphs implies that the BV master equation (\ref{eq:geobvmaster}) is satisfied, which is
due to theorem \ref{thm:1} equivalent to a morphism $\alpha$ from $\oF\oMod(\oCom^{\mathcal{N}=1})$ to $C^{\bullet|\bullet}(\hat{\mathfrak{P}}^{II})$. Second, the factorization properties (\ref{eq:factorizationsym1}), (\ref{eq:factorizationsym2}) and the chain map property qualify a TCFT as a morphism $\beta$ from $C^{\bullet|\bullet}(\hat{\mathfrak{P}}^{II})$ to $\mathcal{E}[A_\alpha,Q_\alpha,\omega_\alpha]$.

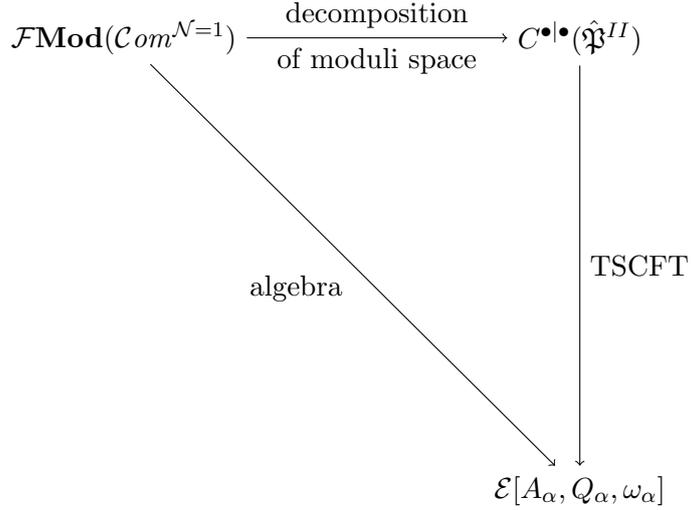
\begin{figure}[h]
\begin{center}
\begin{tikzpicture}[node distance=6cm, auto]
  \node (1) {$\oF\oMod(\oCom^{\mathcal{N}=1})$};
  \node (2) [right of=1] {$C^{\bullet|\bullet}(\hat{\mathfrak{P}}^{II})$};
  \node (3) [below of=2] {$\mathcal{E}[A_\alpha,Q_\alpha,\omega_\alpha]$};
  \draw[->] (1) to node {decomposition} (2);
  \path (1) to node [swap] {of moduli space} (2);
  \draw[->] (2) to node {TSCFT} (3);
  \draw[->] (1) to node [swap]{algebra} (3);
\end{tikzpicture}
\caption{Construction of type II superstring field theory in terms of morphisms of modular operads.}
\label{fig:constructionSFT}
\end{center}
\end{figure}

Schematically, the construction of string field theory can be summarized as depicted in figure \ref{fig:constructionSFT}.

The composition $\gamma\defineL \beta\circ\alpha$ of the morphisms $\alpha$ and $\beta$ then defines an algebra over $\oF\oMod(\oCom^{\mathcal{N}=1})$.
Finally, we want to identify this algebraic structure as some homotopy algebra. We employ the following statements:
\begin{Thm}[\cite{Ginzburg}]
Let $\oP$ be a Koszul cyclic operad. Algebras over the cobar transform (the tree level part of the Feynman transform) of the quadratic dual $\oP^{!}$ of $\oP$ are homotopy $\oP$-algebras.
\end{Thm}
\begin{Def}[\cite{Markl loop}]
Let $\oP$ be a Koszul cyclic operad. Algebras over $\oF\oMod(\oP^{!})$ are loop homotopy $\oP$-algebras.
\end{Def}

Let us first discuss the known results of bosonic string field theory. In closed string field theory, the cyclic operad encoding the symmetry properties of the classical vertices is the operad $\oCom$, whose algebras are commutative algebras. $\oCom$ is Koszul and its quadratic dual is $\oLie$, the operad whose algebras are Lie algebras. A consistent decomposition of the moduli space of closed Riemann surfaces $\mathcal{M}_{g,n}$ defines a morphism from $\oF\oMod(\oCom)$ to $C^\bullet(\hat{\mathcal{P}})$, and a background determines a topological conformal field theory which is a morphism from $C^\bullet(\hat{\mathcal{P}})$ to $\mathcal{E}[A,Q,\omega]$, where $\omega=\op{bpz}(\cdot, c_0^- \cdot)$. Thus the algebraic structure of classical closed string field theory is that of a homotopy Lie-algebra ($L_\infty$-algebra) \cite{Zwiebach closed}, and quantum closed string field theory carries the structure of a loop homotopy Lie-algebra \cite{Markl loop}.

Inspired by that, we call an algebra over $\oF\oMod(\oCom^{\mathcal{N}=1})$ a $\mathcal{N}=1$ loop homotopy Lie-algebra and similarly an algebra over the cobar transform of $\oCom^{\mathcal{N}=1}$ a $\mathcal{N}=1$ homotopy Lie-algebra.

We conclude this section with the following theorem:
\begin{Thm}
The vertices of the quantum/classical master action of type II superstring field theory satisfy the axioms of a $\mathcal{N}=1$ loop homotopy Lie-algebra/$\mathcal{N}=1$ homotopy Lie-algebra.
\end{Thm}

\section{Outlook}
In this paper we outline the construction of type II superstring field theory, leading to a geometric and an algebraic BV master equation analogous to the case in the bosonic string. The construction is based on the small Hilbert space, in contrast to other approaches to superstring field theory like \cite{Berkovits, Zwiebach heterotic}. Picture changing operators arise as the consequence of the fact that we can not define local coordinates around punctures globally on moduli space but just coordinate curves. Pursuing the same idea for classical open superstring field theory would require a restriction of the state space in the Ramond sector, due to the translation invariance in the Ramond divisors. Such a theory might serve as an adequate description of classical open superstring field theory.

Recently, it has been shown that the moduli space of super Riemann surfaces is generically non-split \cite{Witten splitness}. An interesting question is whether the topology of the geometric vertices of type II superstring field theory is considerably simpler than that of the full moduli space, i.e. if the integrals defining the algebraic vertices can be reduced to integrals over the geometric vertices of bosonic closed string field theory.  

A particular feature of type II superstring field theory is, that the symplectic form in the $R-R$ sector degenerates on-shell. On the other hand, this is a necessary condition for a non-trivial open-closed correspondence at the quantum level, as discussed in \cite{classification}. Thus, this indicates that in type I superstring field theory there might be backgrounds where closed strings decouple completely form open strings even at the quantum level, leading to a consistent theory of only open superstrings.

Finally, we describe string field theory in terms of operads. For classical bosonic open strings on a single D-brane, the relevant operad is the operad $\oAss$ of associative algebras. First of all it would be interesting to generalize the operad $\oAss$ to several D-branes, such that algebras over its cobar transform are Calabi-Yau $A_\infty$-categories \cite{Costello}. Second, another project \cite{Martin} is to specify the operad that describes the quantum open-closed homotopy algebra \cite{qocha}.

\vspace{2cm}\noindent {\bf Acknowledgements}: We would like to thank B. Zwiebach, I. Sachs, T. Erler, M. Kroyter and M. Schnabl for many fruitful discussions and comments. Special thanks goes to M. Doubek and M. Markl who had the patience to explain to the authors the theory of operads. K.M. would also like to thank the organizers of the conference ``String Field Theory and Related Aspects V, SFT 2012" hosted by the Israel Institute for Advanced Studies, where part of the work was initiated. The research of B.J. was supported by grant GA\v CR P201/12/G028, whereas K.M. was supported in parts by the DFG Transregional Collaborative Research Centre TRR 33 and the DFG cluster of excellence ``Origin and Structure of the Universe''. We also thank to DAAD (PPP) and ASCR \& MEYS (Mobility) for supporting our collaboration.


\appendix

\section{Super Riemann Surfaces}\label{app:srs}
In this part we follow closely the exposition of \cite{Witten srs}. A super Riemann surface $\Sigma$ is a $1|1$ dimensional complex supermanifold with the additional structure of a subbundle $\D\subset T\Sigma$ of the tangent bundle of rank $0|1$. A Neveu-Schwarz puncture on $\Sigma$ is described by a point $(z,\vt)=(z_0,\vt_0)$, whereas a Ramond puncture is described by a divisor $z=z_0$. The collection of all Ramond punctures defines the Ramond divisor. Note that the number of Ramond punctures is always even.
Furthermore the subbundle $\D$ has to satisfy a non-degeneracy condition: For every non-zero section $D$ of $\D$, the commutator $[D,D]$ has to be linearly independent of $D$ everywhere, except along the Ramond divisor where $[D,D]=0$. Thus a Ramond puncture is part of the structure of a super Riemann surface, in contrast to a  Neveu-Schwarz puncture which merely distinguishes a point on $\Sigma$.
In the following every notion in the Neveu-Schwarz sector will have its counterpart in the Ramond sector, which we will display by NS and R respectively.

A superconformal coordinate system $(z,\vt)$ is distinguished by requiring that every section $D$ of $\D$ is proportional to
\be\no
\Dvt=\pd_\vt + \vt \pd_z \sep \text{NS}
\ee
and
\be\no
\Dvt^\ast=\pd_\vt + z\vt \pd_z \sep \text{R}\com
\ee
where the coordinate system for R covers a subset of $\Sigma$ containing a single R puncture at $z=0$ \footnote{For several R punctures $z_i$, we would have $\Dvt^\ast=\pd_\vt + w(z)\vt \pd_z$ with $w(z)=\Pi_i (z-z_i)$.}.

A superconformal transformation is a change of superconformal coordinates. The general form of such a transformation is
\be\label{eq:confns}
\begin{array}{rcl}
z^\prime &=& u\;\pm\;\vt\alpha \sqrt{u^\prime} \\
\vt^\prime &=& \alpha \;\pm\; \vt  \sqrt{u^\prime} \bracketi{1+\frac{\alpha \alpha^\prime}{2 u^\prime}}
\end{array}
\sep \text{NS}
\ee
and
\be\label{eq:confr}
\begin{array}{rcl}
z^\prime &=& u\;\pm\;\vt\alpha \sqrt{zuu^\prime} \\
\vt^\prime &=& \alpha \;\pm\; \vt  \sqrt{\frac{zu^\prime}{u}} \bracketi{1+\frac{u\alpha \alpha^\prime}{2 u^\prime}}
\end{array}
\sep \text{R}\com
\ee
where $u=u(z)$ is an even function and $\alpha=\alpha(z)$ is odd. The signs in equation (\ref{eq:confns}) and (\ref{eq:confr}) are determined by a choice of branch for the square root of $u^\prime$ and $\tfrac{zu^\prime}{u}$, respectively.

Primary fields of superconformal weight $h$ are defined to be sections of $\D^{-2h}$. Consider for example a function $f\in C^\infty(\Sigma)$, then $D_{\theta} f= (D_{\theta}\theta^\prime)D_{\theta^\prime}f$ transforms as a primary of superconformal weight $1/2$. In general a primary $\phi$ of superconformal weight $h$ can be expanded as $\phi=\varphi_0+\theta \varphi_1$, where $\varphi_0$ has conformal weight $h$ and $\varphi_1$ has conformal weight $h+1/2$.

Finally, a superconformal vectorfield $X$ is a vector field that preserves the subbundle $\D$, that is for every section $D$ of $\D$
\be\no
[X,D]\propto D \pt
\ee
We can choose a basis for the space of superconformal vectorfields which obeys the super Witt algebra:
\be\label{eq:wittns}
\begin{array}{rccc}
l_n &=& -z^{n+1}\pd_z-\inv{2}(n+1)z^n\vt\pd_\vt  \sep & n\in \Z \\
g_n &=& z^{n+1/2}(\pd_\vt-\vt\pd_z) \sep &n\in \Z+1/2
\end{array}
\sep \text{NS}
\ee

\be\label{eq:wittr}
\begin{array}{rccc}
l_n &=& -z^{n+1}\pd_z-\inv{2}nz^n\vt\pd_\vt \sep & n\in \Z \\
g_n &=& z^{n}(\pd_\vt-\vt z\pd_z) \sep & n\in \Z
\end{array}
\sep \text{R}\com
\ee

\be\no
\begin{array}{lcl}
[l_m,l_n] &=& (m-n)l_{m+n} \\
\vspace{0cm}[l_m,g_n] & = & \bracketi{\frac{m}{2}-n}g_{m+n} \\
\vspace{0cm} [g_m,g_n] &=& 2l_{m+n}
\end{array}
\sep \text{NS and R}\pt
\ee

In the NS sector the vectors
\be\no
g_{-1/2}\ssep g_{1/2} \ssep l_{-1} \ssep l_0 \ssep l_1
\ee
form a closed subalgebra and generate the $3|2$ complex dimensional NS M\"{o}bius group $\op{Aut}(S^{2|1}_{NS})$. The general form of a NS M\"{o}bius transformation is given by
\be\label{eq:mobns}
\bpm
z^\prime\\ \vt^\prime
\epm
=
\bpm
\displaystyle
\frac{az+b}{cz+d} &\pm& \displaystyle \vt \frac{\gamma z+\delta}{(cz+d)^2}   \vspace{0.1cm} \\ \vspace{0.1cm}
\displaystyle
\frac{\gamma z+\delta}{cz+d} &\pm& \displaystyle \vt \frac{1+\ttinv{2}\delta\gamma}{cz+d}
\epm
\com
\ee
where $a,b,c,d\in\C^{1|0}$, $\gamma,\delta\in \C^{0|1}$ and $ad-bc=1$.

The maximal non-trivial subalgebra in the R sector is spanned by
\be\no
l_{-1} \ssep l_0 \ssep l_1
\ee
and generates the $3|0$ complex dimensional R M\"{o}bius group $\op{Aut}(S^{2|1}_{R})$. A generic element of $\op{Aut}(S^{2|1}_{R})$ takes the form
\be\label{eq:mobr}
\bpm
z^\prime\\ \vt^\prime
\epm
=
\bpm
\displaystyle
\frac{az+b}{cz+d} \vspace{0.1cm} \\ \vspace{0.1cm}
\displaystyle
\pm\vt \bracketii{\frac{z}{(az+b)(cz+d)}}^{1/2}
\epm
\com
\ee
with $a,b,c,d\in\C^{1|0}$ and $ad-bc=1$.

\section{Superconformal Field Theory of Type II String}\label{app:scft}
The field content of the superconformal field theory of type II string theory is composed of matter fields and ghost fields.
The matter sector is described by scalars
\be\no
X^\mu(z,\ti{z},\vt,\ti{\vt}) \com
\ee
and the ghost sector contains the holomorphic ghosts
\be\no
B= \beta + \vt b \mand C =c+\vt \gamma \com
\ee
of superconformal weight $(3/2,0)$ and $(-1,0)$, respectively, and the antiholomorphic ghosts
\be\no
\ti{B}= \ti{\beta} + \ti{\vt}  \ti{b} \mand  \ti{C} = \ti{c}+  \ti{\vt}\ti{\gamma} \com
\ee
of superconformal weight $(0,3/2)$ and $(0,-1)$, respectively.
Let $\phi$ be a holomorphic local operator of superconformal weight $h$ in the NS sector and $(z,\vt)=(z_{\op{radial}},\vt_{\op{radial}})$ the standard coordinate system of radial quantization, then the mode expansion of $\phi$ reads
\be\label{eq:modens}
\phi(z,\vt)=\sum_{n\in \Z} \frac{\phi^0_n}{z^{n+h}}+\vt\sum_{n\in \Z+1/2} \frac{\phi^1_n}{z^{n+h+1/2}} \pt
\ee
Now consider a holomorphic local operator $\phi$ of superconformal weight $h$ in the R sector. The coordinate system of radial quantization is not a good coordinate system in the R sector - it involves a branch cut \cite{Alvarez}. We obtain a superconformal coordinate system in the sense of (\ref{eq:confr}) by defining new coordinates $(z,\vt)=(z_{\op{radial}},\theta_{\op{radial}} z_{\op{radial}}^{-1/2})$.
In these coordinates, the mode expansion reads
\be\label{eq:moder}
\phi(z,\vt)=\sum_{n\in \Z} \frac{\phi^0_n}{z^{n}}+\vt\sum_{n\in \Z } \frac{\phi^1_n}{z^{n}}\pt
\ee
The sewing maps (\ref{eq:bpznspm}) and (\ref{eq:bpzrpm}) define the bpz conjugation
\be\no
\op{bpz}_{NS}(\phi)(z,\vt)=(I_{(+,+)}^\ast\phi)(z,\vt) \mand \op{bpz}_{R}(\phi)(z,\vt)=(I_{(+,-)}^\ast\phi)(z,\vt)\pt
\ee
From the mode expansion (\ref{eq:modens}) and (\ref{eq:moder}) we can infer that
\begin{align}\no
\op{bpz}_{NS}(\phi^0_n)=(-1)^{n+h}\phi^0_{-n} \sep \op{bpz}_{NS}(\phi^1_n)=(-1)^{n+h+1/2}\phi^1_{-n} \com \\
\op{bpz}_{R}(\phi^0_n)=(-1)^{n+h}\phi^0_{-n} \sep \op{bpz}_{R}(\phi^1_n)=(-1)^{n+h+1/2}\phi^1_{-n} \com
\end{align}
which is indeed the same for every sector and every type of mode.

The operator state correspondence is formulated in terms of the coordinates of radial quantization, so there is no problem in the NS sector. In the R sector, in contrast, the coordinates of radial quantization are ill defined. To resolve this problem, one introduces spin fields which map the NS ground state to the R ground state \cite{FMS}. We denote the spin fields in the matter sector by
\be\no
S_m^{s_1,\dots,s_5}(z) \com
\ee
and in the ghost sector by
\be\no
S_g^{\pm}(z)\com
\ee
such that
\be\no
S_g^- S_m^{s_1,\dots,s_5}\ket{0}_{NS}=\ket{s_1,\dots,s_5}_R
\ee
describes the R ground state.
Furthermore, the ghost spin field satisfies \cite{FMS, Sen spinfield}
\begin{align}\label{eq:opesf}
\beta(z_1)S_g^{\pm}(z_2)\sim z_{12}^{\pm1/2} :\beta S_{g}^{\pm}:(z_2) \\\no
\gamma(z_1)S_g^{\pm}(z_2)\sim z_{12}^{\mp1/2} :\gamma S_{g}^{\pm}:(z_2) \pt
\end{align}

The operator state correspondence together with (\ref{eq:opesf}) determines the creation operators in the ghost sector to be
\begin{align}\no
&\dots, \gamma_{-1/2},\gamma_{1/2}\\\no
&\dots,\beta_{-5/2},\beta_{-3/2}\\\no
&\dots, c_{0},c_{1}\\\no
&\dots,b_{-3},b_{-2}
\end{align}
in the NS sector and
\begin{align}\no
&\dots, \gamma_{-1},\gamma_{0}\\\no
&\dots,\beta_{-2},\beta_{-1}\\\no
&\dots, c_{0},c_{1}\\\no
&\dots,b_{-3},b_{-2}
\end{align}
in the R sector.
The creation operators whose bpz conjugate is also a creation operator are called zero modes. This determines the ghost zero modes
\begin{align}\no
\gamma_{-1/2}&,\gamma_{1/2}\\\no
c_{-1},&c_0,c_1
\end{align}
in the NS sector and
\begin{align}\no
&\gamma_0\\\no
c_{-1},&c_0,c_1
\end{align}
in the R sector.
In order to obtain a non-vanishing correlator, one has to saturate these zero modes. This requires an insertion
\be\no
c_{-1}c_0c_1 \delta(\gamma_{-1/2})\delta(\gamma_{1/2})
\ee
and
\be\no
c_{-1}c_0c_1 \delta(\gamma_0)
\ee
in the NS and R sector, respectively. A geometric interpretation of delta functions of ghost operators has first been given in \cite{Belopolsky geometric, Belopolsky picture}, which we review in appendix \ref{app:forms} together with the rules how to manipulate such expressions. Furthermore, the geometric interpretation suggests a grading which differs from the conventional ghost number and picture grading:
We define ghost number by assigning ghost number one to $c,\gamma$ and ghost number minus one to $b,\beta$. Picture number is associated with delta functions of Grassmann even ghosts, that is
$\delta(\gamma_n)$ carries picture number one and $\delta(\beta_n)$ carries picture number minus one. Finally, we set the ghost number and picture for both the NS and the R groundstate to be zero. We will denote ghost number and picture collectively by $g|p$.

Thus
\be\no
\op{deg}(\ket{0}_{NS})=0|0 \sep \op{deg(}\ket{s_1,\dots,s_5}_R)=0|0
\ee
implies
\be\no
\op{deg}(S^-_g)=0|0\pt
\ee
The bpz inner product of states $\Phi_1$ and $\Phi_2$ is defined by
\be\label{eq:bpzalg}
\op{bpz}_\alpha(\Phi_1,\Phi_2)\defineL \left\langle (I_\alpha^\ast\Phi_1) \Phi_2 \right\rangle \com
\ee
where $I_\alpha$ is the sewing map defined in (\ref{eq:bpzgeo}). Thus we conclude that
\begin{align}\no
\op{deg}\bracketi{\op{bpz}_{NS-NS}}&= -6|-4\\\no
\op{deg}\bracketi{\op{bpz}_{R-NS}}= \op{deg}\bracketi{\op{bpz}_{NS-R}}&=-6|-3\\\no
\op{deg}\bracketi{\op{bpz}_{R-R}}&=-6|-2 \com
\end{align}
Moreover we have
\be\no
S^-_g(z_1)S^-_g(z_2)\sim \inv{z_{12}^{1/4}} \delta(\gamma)(z_2) \com
\ee
which implies that the OPE of two R vertex operators carries degree $0|1$.

To proceed, we depict maps on the state space of the CFT by directed graphs, where the direction which distinguishes inputs and outputs points from left to right. Thus, the bpz inner product in the corresponding sectors is represented by\\
\bt
\path (0,0.5cm) node[shape=circle,anchor=east] (i1) {$\scriptscriptstyle ++$};
\path (0,-0.5cm) node[shape=circle,anchor=east] (i2) {$\scriptscriptstyle ++$};
\path (1cm,0) node[shape=circle] () {$\com $};
\path (0,-1cm) node[shape=circle] () {$\scriptstyle-6|-4$};
\draw[thick] (0,-0.5cm) arc (-90:90:0.5cm);
\et
\bt
\path (0,0.5cm) node[shape=circle,anchor=east] (i1) {$\scriptscriptstyle +-$};
\path (0,-0.5cm) node[shape=circle,anchor=east] (i2) {$\scriptscriptstyle +-$};
\path (1cm,0) node[shape=circle] () {$\com $};
\path (0,-1cm) node[shape=circle] () {$\scriptstyle-6|-3$};
\draw[thick] (0,-0.5cm) arc (-90:90:0.5cm);
\et
\bt
\path (0,0.5cm) node[shape=circle,anchor=east] (i1) {$\scriptscriptstyle -+$};
\path (0,-0.5cm) node[shape=circle,anchor=east] (i2) {$\scriptscriptstyle -+$};
\path (1cm,0) node[shape=circle] () {$\com $};
\path (0,-1cm) node[shape=circle] () {$\scriptstyle-6|-3$};
\draw[thick] (0,-0.5cm) arc (-90:90:0.5cm);
\et
\bt
\path (0,0.5cm) node[shape=circle,anchor=east] (i1) {$\scriptscriptstyle --$};
\path (0,-0.5cm) node[shape=circle,anchor=east] (i2) {$\scriptscriptstyle --$};
\draw[thick] (0,-0.5cm) arc (-90:90:0.5cm);
\path (0,-1cm) node[shape=circle] () {$\scriptstyle-6|-2$};
\et
\\ and its inverse by\\
\bt
\path (0,0.5cm) node[shape=circle,anchor=west] (i1) {$\scriptscriptstyle ++$};
\path (0,-0.5cm) node[shape=circle,anchor=west] (i2) {$\scriptscriptstyle ++$};
\path (1.2cm,0) node[shape=circle] () {$\com $};
\path (0,-1cm) node[shape=circle] () {$\scriptstyle6|4$};
\draw[thick] (0,-0.5cm) arc (270:90:0.5cm);
\et
\bt
\path (0,0.5cm) node[shape=circle,anchor=west] (i1) {$\scriptscriptstyle +-$};
\path (0,-0.5cm) node[shape=circle,anchor=west] (i2) {$\scriptscriptstyle +-$};
\path (1.2cm,0) node[shape=circle] () {$\com $};
\path (0,-1cm) node[shape=circle] () {$\scriptstyle6|3$};
\draw[thick] (0,-0.5cm) arc (270:90:0.5cm);
\et
\bt
\path (0,0.5cm) node[shape=circle,anchor=west] (i1) {$\scriptscriptstyle -+$};
\path (0,-0.5cm) node[shape=circle,anchor=west] (i2) {$\scriptscriptstyle -+$};
\path (1.2cm,0) node[shape=circle] () {$\com $};
\path (0,-1cm) node[shape=circle] () {$\scriptstyle6|3$};
\draw[thick] (0,-0.5cm) arc (270:90:0.5cm);
\et
\bt
\path (0,0.5cm) node[shape=circle,anchor=west] (i1) {$\scriptscriptstyle --$};
\path (0,-0.5cm) node[shape=circle,anchor=west] (i2) {$\scriptscriptstyle --$};
\path (1.2cm,0) node[shape=circle] () {$\com $};
\path (0,-1cm) node[shape=circle] () {$\scriptstyle6|2$};
\draw[thick] (0,-0.5cm) arc (270:90:0.5cm);
\et
\\where we abbreviate NS and R as $+$ and $-$, respectively, and also indicate the degree. Similarly, the OPE is depicted by\\
\bt
\path (0:0.9cm) node[anchor=west] (o) {$\scriptscriptstyle ++$};
\path (120:0.9cm) node[anchor=east] (i1) {$\scriptscriptstyle ++$};
\path (240:0.9cm) node[anchor=east] (i2) {$\scriptscriptstyle ++$};
\path (1.7cm,0) node[shape=circle] () {$\com $};
\path (0.3cm,-1.2cm) node[shape=circle] () {$\scriptstyle0|0$};
\path (0,0) coordinate (0);
\draw[thick] (0) -- (i1) (0) -- (i2) (0) -- (o);
\et
\bt
\path (0:0.9cm) node[anchor=west] (o) {$\scriptscriptstyle ++$};
\path (120:0.9cm) node[anchor=east] (i1) {$\scriptscriptstyle -+$};
\path (240:0.9cm) node[anchor=east] (i2) {$\scriptscriptstyle -+$};
\path (1.7cm,0) node[shape=circle] () {$\com $};
\path (0.3cm,-1.2cm) node[shape=circle] () {$\scriptstyle0|1$};
\path (0,0) coordinate (0);
\draw[thick] (0) -- (i1) (0) -- (i2) (0) -- (o);
\et
\bt
\path (0:0.9cm) node[anchor=west] (o) {$\scriptscriptstyle ++$};
\path (120:0.9cm) node[anchor=east] (i1) {$\scriptscriptstyle +-$};
\path (240:0.9cm) node[anchor=east] (i2) {$\scriptscriptstyle +-$};
\path (1.7cm,0) node[shape=circle] () {$\com $};
\path (0.3cm,-1.2cm) node[shape=circle] () {$\scriptstyle0|1$};
\path (0,0) coordinate (0);
\draw[thick] (0) -- (i1) (0) -- (i2) (0) -- (o);
\et
\bt
\path (0:0.9cm) node[anchor=west] (o) {$\scriptscriptstyle ++$};
\path (120:0.9cm) node[anchor=east] (i1) {$\scriptscriptstyle --$};
\path (240:0.9cm) node[anchor=east] (i2) {$\scriptscriptstyle --$};
\path (1.7cm,0) node[shape=circle] () {$\com $};
\path (0.3cm,-1.2cm) node[shape=circle] () {$\scriptstyle0|2$};
\path (0,0) coordinate (0);
\draw[thick] (0) -- (i1) (0) -- (i2) (0) -- (o);
\et
\bt
\path (0:0.9cm) node[anchor=west] (o) {$\scriptscriptstyle -+$};
\path (120:0.9cm) node[anchor=east] (i1) {$\scriptscriptstyle ++$};
\path (240:0.9cm) node[anchor=east] (i2) {$\scriptscriptstyle -+$};
\path (1.7cm,0) node[shape=circle] () {$\com $};
\path (0.3cm,-1.2cm) node[shape=circle] () {$\scriptstyle0|0$};
\path (0,0) coordinate (0);
\draw[thick] (0) -- (i1) (0) -- (i2) (0) -- (o);
\et
\vspace{0.5cm}\\
\bt
\path (0:0.9cm) node[anchor=west] (o) {$\scriptscriptstyle +-$};
\path (120:0.9cm) node[anchor=east] (i1) {$\scriptscriptstyle ++$};
\path (240:0.9cm) node[anchor=east] (i2) {$\scriptscriptstyle +-$};
\path (1.7cm,0) node[shape=circle] () {$\com $};
\path (0.3cm,-1.2cm) node[shape=circle] () {$\scriptstyle0|0$};
\path (0,0) coordinate (0);
\draw[thick] (0) -- (i1) (0) -- (i2) (0) -- (o);
\et
\bt
\path (0:0.9cm) node[anchor=west] (o) {$\scriptscriptstyle --$};
\path (120:0.9cm) node[anchor=east] (i1) {$\scriptscriptstyle ++$};
\path (240:0.9cm) node[anchor=east] (i2) {$\scriptscriptstyle --$};
\path (1.7cm,0) node[shape=circle] () {$\com $};
\path (0.3cm,-1.2cm) node[shape=circle] () {$\scriptstyle0|0$};
\path (0,0) coordinate (0);
\draw[thick] (0) -- (i1) (0) -- (i2) (0) -- (o);
\et
\bt
\path (0:0.9cm) node[anchor=west] (o) {$\scriptscriptstyle --$};
\path (120:0.9cm) node[anchor=east] (i1) {$\scriptscriptstyle -+$};
\path (240:0.9cm) node[anchor=east] (i2) {$\scriptscriptstyle +-$};
\path (1.7cm,0) node[shape=circle] () {$\com $};
\path (0.3cm,-1.2cm) node[shape=circle] () {$\scriptstyle0|0$};
\path (0,0) coordinate (0);
\draw[thick] (0) -- (i1) (0) -- (i2) (0) -- (o);
\et
\bt
\path (0:0.9cm) node[anchor=west] (o) {$\scriptscriptstyle +-$};
\path (120:0.9cm) node[anchor=east] (i1) {$\scriptscriptstyle -+$};
\path (240:0.9cm) node[anchor=east] (i2) {$\scriptscriptstyle --$};
\path (1.7cm,0) node[shape=circle] () {$\com $};
\path (0.3cm,-1.2cm) node[shape=circle] () {$\scriptstyle0|1$};
\path (0,0) coordinate (0);
\draw[thick] (0) -- (i1) (0) -- (i2) (0) -- (o);
\et
\bt
\path (0:0.9cm) node[anchor=west] (o) {$\scriptscriptstyle -+$};
\path (120:0.9cm) node[anchor=east] (i1) {$\scriptscriptstyle +-$};
\path (240:0.9cm) node[anchor=east] (i2) {$\scriptscriptstyle --$};
\path (1.7cm,0) node[shape=circle] () {$\pt $};
\path (0.3cm,-1.2cm) node[shape=circle] () {$\scriptstyle0|1$};
\path (0,0) coordinate (0);
\draw[thick] (0) -- (i1) (0) -- (i2) (0) -- (o);
\et
\\Now one can construct arbitrary surfaces from these elementary ones, and thus determine the degree of a correlation function $Z({\bf\Sigma}_{g,\vect{n}})$ on a type II world sheet ${\bf\Sigma}_{g,\vect{n}}$ to be
\be\label{eq:degZ}
\op{deg}\bracketi{Z({\bf\Sigma}_{g,\vect{n}})}=6g-6|4g-4+ n_{R-R} +\inv{2}(n_{R-NS}+n_{NS-R}) \pt
\ee

Finally, the typical form of a vertex operator is given by
\begin{align}\label{eq:goundstates}
c\delta(\gamma)\,\ti{c}\delta(\ti{\gamma})\, V \sep &2|2 \sep \text{NS-NS} \com \\\no
c\delta(\gamma)\,\ti{c}\ti{S}^-_g\ti{S}^{\ti{s}_1,\dots,\ti{s}_5}_m\,V\sep &2|1 \sep \text{NS-R} \com \\\no
{c}{S}^-_g{S}^{s_1,\dots,s_5}_m\,\ti{c}\delta(\ti{\gamma})\, V \sep &2|1 \sep \text{NS-R} \com \\\no
{c}{S}^-_g{S}^{s_1,\dots,s_5}_m\,\ti{c}\ti{S}^-_g\ti{S}^{\ti{s}_1,\dots,\ti{s}_5}_m\,V \sep &2|0 \sep \text{R-R} \com
\end{align}
where $V$ represents some matter vertex operator.

\section{Forms in Supergeometry and Relation to String Theory}\label{app:forms}
The superconformal ghosts of superstring theory can be interpreted as operations acting on differential forms \cite{Belopolsky picture, Belopolsky geometric}. To illustrate this analogy, we will start with a brief review of geometric integration theory on supermanifolds \cite{Voronov, Witten integration, Belopolsky picture, Belopolsky geometric}.

Let $M^{m|n}$ be a $m|n$ dimensional supermanifold. A differential form $\omega\in \Omega^{r|s}(M^{m|n})$ is a function of $r$ even and $s$ odd tangent vectors, which satisfies
\be\label{eq:symforms}
\omega(g\B{V})=\op{sdet}(g) \omega(\B{V}) \sep \forall g\in \op{GL}(r|s)
\ee
and
\be\no
\bracket{\pd_{V_A^M}\pd_{V_B^N}-(-1)^{AB+N(A+B)}\pd_{V_B^M}\pd_{V_A^N}}\omega(\B{V})=0 \com
\ee
where $\B{V}=(v_1,\dots,v_r|\nu_1,\dots,\nu_s)$ denotes a collection of tangent vectors and $V_A^M$ is the $M$-th component of the $A$-th tangent vector, i.e. $A,B\in \{1,\dots,r|s\}$ and $M,N\in \{1,\dots,m|n\}$. The exterior derivative $d:\omega^{r|s}(M^{m|n})\to \omega^{r+1|s}(M^{m|n})$ is defined by
\be\no
(d\omega)(v_1,\dots,v_r,v_{r+1},\nu_1,\dots,\nu_s)=(-1)^rv^M_{r+1}(\delta_{x^M}\,\omega)(v_1,\dots,v_r,\nu_1,\dots,\nu_s) \com
\ee
where
\be\no
(\delta_{x^M}\,\omega)(\B{V})=\pd_{x^M}\omega(\B{V})-(-1)^{MA}V^N_A\pd_{x^N}\pd_{V^M_A}\omega(\B{V})\com
\ee
and $x^M$ are coordinates on $M^{m|n}$. Let $V$ be a vector field on $M^{m|n}$. The interior product $i_V:\omega^{r|s}(M^{m|n})\to \omega^{r-1|s}(M^{m|n})$ is defined by
\be\no
(i_V\omega)(v_1,\dots,v_{r-1}|\nu_1,\dots,\nu_s)=\omega(V,v_1,\dots,v_{r-1}|\nu_1,\dots,\nu_s) \pt
\ee
The space of differential forms is preserved under multiplication with functions. Thus, imposing the Leibniz rule w.r.t. $d$ makes $\Omega^{r|s}(M^{m|n})$ a module over $\Omega^{r^\prime|0}(M^{m|n})$, in particular over $1$-forms. We denote the operation of multiplying a $1$-form $\alpha\in\Omega^{1|0}(M^{m|n})$ by $e_\alpha:\Omega^{r|s}(M^{m|n})\to\Omega^{r+1|s}(M^{m|n})$, which explicitly reads
\be\no
(e_\alpha \omega)(v_1,\dots,v_r,v_{r+1}|\nu_1,\dots,\nu_s)=(-1)^r \bracketii{\alpha(v_{r+1})\omega(\B{V}) -(-1)^{MA}\alpha(V_A)V_{r+1}^M\pd_{V^M_A}\omega(\B{V})} \pt
\ee
The operations introduced so far just affect the number of even vectors, but there are also operations witch change the number of odd vectors: Let $\nu$ be an odd vector field on $M^{m|n}$. The operation
$\delta(i_\nu):\Omega^{r|s}(M^{m|n})\to\Omega^{r|s-1}(M^{m|n})$ is defined by
\be\label{eq:deltai}
\bracketi{\delta(i_\nu)\omega}(v_1,\dots,v_r|\nu_1,\dots,\nu_{s-1})=\omega(v_1,\dots,v_r|\nu,\nu_1,\dots,\nu_{s-1}).
\footnote{In (\ref{eq:deltai}), we use a different sign convention than in the original work \cite{Belopolsky picture}, which is more natural in the context of superstring theory.}
\ee
Similarly, for an odd $1$-form $\beta$, there is an operation $\delta(e_\beta):\Omega^{r|s}(M^{m|n})\to\Omega^{r|s+1}(M^{m|n})$,
\be\label{eq:deltae}
\delta(e_\beta)(v_1,\dots,v_r|\nu_1,\dots,\nu_s,\nu_{s+1})=\inv{\beta(\nu_{s+1})}\omega\bracketii{\dots,V_A-\frac{\beta(V_A)}{\beta(\nu_{s+1})}\nu_{s+1},\dots} \pt
\ee
The grading is defined by $r|s$ plus the Grasssmann parity $p\in \Z_2$, which we denote collectively by $r|s|p$. To summarize, we have five basic operations on the space of differential forms, listed in table \ref{tab:basicop} together with the corresponding degrees.

\begin{table}
\begin{center}
\begin{tabular}{c|c|c|c|c}
$d$ & $i_V$ & $e_\alpha$ & $\delta(i_\nu)$ & $\delta(e_\beta)$  \\\hline
$\quad1|0|0\quad$ & $\quad-1|0|V\quad$ & $\quad1|0|\alpha\quad$ & $\quad0|-1|1\quad$ & $\quad0|1|1\quad$
\end{tabular}\caption{Basic operations on differential forms. The Grassmann parity of $V$ and $\alpha$ is undetermined, whereas $\beta$ and $\nu$ are odd.}\label{tab:basicop}
\end{center}
\end{table}

Differential $r|s$-forms on $M^{m|n}$ are the natural objects for integrating $r|s$ dimensional submanifolds of $M^{m|n}$. But as in the even case, one needs an orientation on the submanifold to carry out the integration unambiguously. The general linear group $\op{GL}(m|n)$ has four normal subgroups, which determine the possible notions of orientability:
\bi
\item[(i)] $[++]$ orientation: $\op{det}(g_{00})>0$ and $\op{det}(g_{11})>0$
\item[(ii)] $[+-]$ orientation: $\op{det}(g_{00})>0$
\item[(iii)] $[-+]$ orientation: $\op{det}(g_{11})>0$
\item[(iv)] $[--]$ orientation: $\op{det}(g_{00})\op{det}(g_{11})>0$
\ei
where
\be\no
\op{GL}(m|n)\ni g=\bpm g_{00} & g_{01}\\ g_{10} & g_{11} \epm \pt
\ee
Due to the symmetry properties of differential forms (\ref{eq:symforms}) and the fact that $\int d\theta_1d\theta_2=-\int d\theta_2d\theta_1$ whereas $\int dx_1dx_2=\int dx_2dx_1$, the appropriate orientation for integrating differential forms is the $[+-]$ orientation.

Let $A_1$ and $A_2$ be some operators on the space of differential forms of degree $r_1|s_1|p_1$ and $r_2|s_2|p_2$, respectively. We define the commutator to be
\be\no
[A_1,A_2]=A_1\circ A_2 -(-1)^{A_1A_2}A_2\circ A_1 \com
\ee
 where
 \be\label{eq:degsign}
 (-1)^{A_1A_2}=(-1)^{r_1r_2+p_1p_2} \pt
 \ee
Note that $s_1$ and $s_2$ do not occur in equation (\ref{eq:degsign}), which is in accordance with the $[+-]$ orientation. Thus the part $s$ of the grading does not produce a sign upon permutation, as it has been already observed in section \ref{sec:geobv} in the context of the oriented singular chain complex of moduli spaces.

 In the following we describe some operations generated from the basic operations of table \ref{tab:basicop}. The Lie derivative w.r.t. a vector field $V$ is defined by
 \be\no
 \mathcal{L}_V=[d,i_V] \sep \op{deg}( \mathcal{L}_V)=0|0|V \pt
 \ee
Furthermore
\be\no
[e_\beta,\delta(i_\nu)]= -\beta(\nu)\delta^\prime(i_\nu) \sep
\ee
and more generally
\be\no
[e_\beta,\delta^{(n)}(i_\nu)]= -\beta(\nu)\delta^{(n+1)}(i_\nu) \sep \op{deg}(\delta^{(n)}(i_\nu))=n|-1|n+1 \pt
\ee
Similarly
\be\no
[i_\nu,\delta^{(n)}(e_\beta)]= \beta(\nu)\delta^{(n+1)}(e_\beta) \sep \op{deg}(\delta^{(n)}(e_\beta))=-n|1|n+1 \pt
\ee
The picture changing operator $\Gamma_\nu$ of degree $0|1|0$ associated to an odd vector field $\nu$ is defined by \cite{Belopolsky picture}
\begin{align}\label{eq:pcogeo}
\Gamma_\nu&=\inv{2}\bracketi{\mathcal{L}_\nu\,\delta(i_\nu)-\delta(i_\nu)\,\mathcal{L}_\nu}\\\no
&=\mathcal{L}_\nu\,\delta(i_\nu)+\inv{2}i_{[\nu,\nu]}\,\delta^\prime(i_\nu)\\\no
&=-\delta(i_\nu)\,\mathcal{L}_\nu -\inv{2}i_{[\nu,\nu]}\delta^\prime(i_\nu)
\end{align}
where the second and the third line of equation (\ref{eq:pcogeo}) are derived by using relations of (\ref{eq:comrelations}).

The following identities hold:
\begin{align}\label{eq:comrelations}
[e_\alpha,i_V]&=\alpha(V)\, \op{id}\\\no
[\mathcal{L}_{V_1},i_{V_2}]&=i_{[V_1,V_2]} \\\no
[\mathcal{L}_{V_1},\mathcal{L}_{V_2}]&=\mathcal{L}_{[V_1V_2]} \\\no
[\delta(i_{\nu_1}),\delta(i_{\nu_2})]=[\delta(e_{\alpha_1}),\delta(e_{\alpha_2})]&=0 \\\no
[\delta(i_\nu),i_V]=[\delta(e_\beta),e_\alpha]&=0 \\\no
[i_{V_1},i_{V_2}]=[e_{\alpha_1},e_{\alpha_2}]&=0 \\\no
[d,\mathcal{L}_{V}]&=0 \\\no
[d,\Gamma_{\nu}]&=0 \\\no
[\mathcal{L}_{\nu},\delta^{(n)}(i_\nu)]&=-i_{[\nu,\nu]}\delta^{(n+1)}(i_\nu) \\\no
[d,\delta^{(n)}(i_\nu)]&=-\mathcal{L}_\nu \delta^{(n+1)}(i_\nu)-\inv{2}i_{[\nu,\nu]}\delta^{(n+2)}(i_\nu) \pt
\end{align}
Finally,
\be\no
\delta(i_\nu)\delta(e_\beta)=-\inv{\beta(\nu)} \sep \text{on forms annihilated by $i_\nu$} \com
\ee
and similarly
\be\no
\delta(e_\beta)\delta(i_\nu)=\inv{\beta(\nu)} \sep \text{on forms annihilated by $e_\beta$} \pt
\ee

The relation to superstring theory is established by the identifications \cite{Belopolsky picture}
\begin{align}\label{eq:identifications}
b_n &\leftrightarrow i_{l_n} \\\no
c_n &\leftrightarrow e_{l^\ast_{-n}} \\\no
\beta_n &\leftrightarrow i_{g_n} \\\no
\gamma_n &\leftrightarrow e_{g^\ast_{-n}} \\\no
B(V) &\leftrightarrow i_V \\\no
T(V) &\leftrightarrow \mathcal{L}_V \\\no
X_\nu &\leftrightarrow \Gamma_\nu \\\no
Q &\leftrightarrow d \com
\end{align}
where $\{l^\ast_n\}$ and $\{g^\ast_n\}$ represents the dual basis of $\{l_n\}$ and $\{g_n\}$, i.e. $l^\ast_m(l_n)=\delta_{m,n}$ and $g^\ast_m(g_n)=\delta_{m,n}$. The identities (\ref{eq:comrelations}) hold also with the replacements of (\ref{eq:identifications}). The grading in the superconformal field theory is traditionally denoted by $g|p|\alpha$, rather than $r|s|p$, referring to ghost number, picture and Grassmann parity respectively.


\end{document}